\documentstyle[aps,prb,twocolumn,floats,amsfonts,epsf]{revtex}

\tighten
\draft

\begin{document}
\newcommand{\BE}{\begin{equation}}
\newcommand{\EE}{\end{equation}}
\newcommand{\BA}{\begin{eqnarray}}
\newcommand{\EA}{\end{eqnarray}}
\newcommand{\BAA}{\begin{eqnarray*}}
\newcommand{\EAA}{\end{eqnarray*}}
\newcommand{\D}{\displaystyle}
\renewcommand{\arraystretch}{1.5}
\renewcommand{\bottomfraction}{0.9}
\renewcommand{\topfraction}{0.9}

\title{Surface Critical Behavior of bcc Binary Alloys}
\author{R.\ Leidl and H.~W.\ Diehl}
\address{Fachbereich Physik, Universit\"at-Gesamthochschule Essen,\\
D-45117 Essen, Federal Republic of Germany}
\date{July 25, 1997}
\maketitle
\begin{abstract}
The surface critical behavior of bcc binary alloys
undergoing a continuous $B2$-$A2$ order-disorder transition in the bulk
is investigated in the mean-field (MF) approximation, employing a semi-infinite
lattice model equivalent to an Ising antiferromagnet in an external field.
Our main aim is to present clear evidence for the fact that
surfaces which {\em break the two-sublattice symmetry\/} generically display
the critical behavior of the {\em normal\/} transition, whereas
symmetry-preserving surfaces exhibit the behavior of the ordinary transition.
To this end, the lattice MF equations
for both symmetry-breaking (100) and symmetry-preserving (110) surfaces
are cast in the form of nonlinear symplectic maps,
the associated Hamiltonian flows are analyzed, and the length scales involved
are computed. Careful examination of the continuum limit yields
the appropriate semi-infinite Ginzburg-Landau model for the (100) surface
and reveals subtleties overlooked in previous work.
The continuum model involves an ``effective'' {\em ordering surface field\/}
$g_1\neq0$, which depends on the parameters of the lattice model.
The singular behavior predicted by the Ginzburg-Landau model is shown to agree
quantitatively with the solutions of the lattice MF equations.
\end{abstract}
\narrowtext


\section{Introduction}\label{Intro}
Experiments on binary ($AB$) alloys that undergo an order-disorder
transition in the bulk have yielded a wealth of information
on surface critical phenomena in semi-infinite matter.\cite{Dosch}
In these systems one inevitably has to cope with the influence
of {\em surface segregation}, i.~e., the enrichment of one component
at the surface. Surface segregation occurs, e.~g.,
due to different interaction energies or sizes of the two species.
Theoretically, the variation of the local composition near the surface
may necessitate the introduction of ``nonordering'' densities, which are given
by linear combinations of the local concentrations of $A$ and $B$ atoms
on the various sublattices. In the case of surface critical phenomena
at {\em first-order\/} bulk transitions,
such as ``surface-induced disordering'' in fcc binary alloys,
nonordering densities strongly influence the asymptotic behavior.\cite{KG}
In this paper we are concerned with bcc alloys that exhibit
a {\em continuous\/} (second-order) bulk transition and are thus
promising candidates for testing current theories of surface critical behavior
at bulk critical points.\cite{Binder-DL8,Diehl-DL10,Diehl-RG96}
\begin{figure}[b]
\epsfxsize=8.6cm
\epsffile{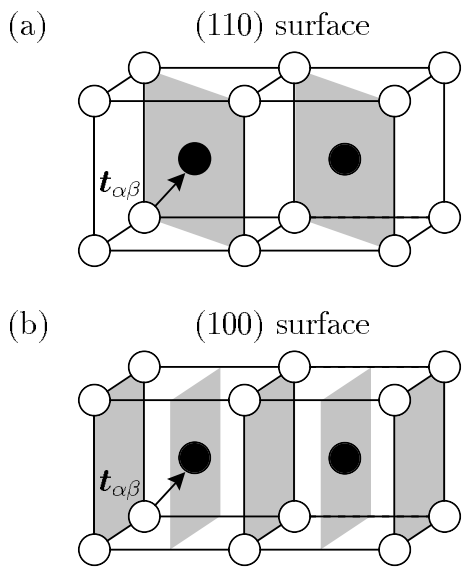}
\vspace{2mm}
\caption
{
Examples of (a) {\em symmetry-preserving}, and (b) {\em symmetry-breaking\/}
surfaces. Sites of sublattices $\alpha$ and $\beta$
correspond to open and full circles, respectively. In the $A2$ phase,
the concentration of either component is the same on all sites,
whereas the two sublattices are preferentially occupied
by $A$ and $B$ atoms, respectively, in the ordered $B2$ phase.
}
\label{Fig_B2}
\end{figure}

The continuous $B2$-$A2$ transition occurring in FeAl or FeCo
has been investigated previously by F.~Schmid.\cite{Schmid}
She studied a semi-infinite lattice model equivalent to a bcc Ising
antiferromagnet both by Monte Carlo simulation and within the mean-field
(MF) approximation, and made the important observation that the orientation
of the surface in general matters. Her conclusions can be summarized
as follows: (a) A nonvanishing order parameter (OP) profile occurs
for $T\ge T_c$, the bulk critical temperature, provided that (i) the surface
breaks the two-sublattice symmetry (see below), and (ii) one component
is enriched at the surface.
(b$^\prime$) The observable surface critical behavior should be representative
of the {\em ordinary\/} universality class even if the above conditions
(i) and (ii) are met.

While we agree with (a), we find that (b$^\prime$) should be replaced by (b):
If conditions (i) and (ii) are satisfied, the surface critical behavior
generically is characteristic of the {\em normal\/} transition,
which belongs to the same universality class as the {\em extraordinary\/}
transition.\cite{normal}

In a foregoing Letter \cite{Drewitz} by Drewitz, Burkhardt, and ourselves,
exact transfer matrix calculations were employed in conjunction with conformal
invariance to present clear evidence for (a) and (b) in bulk dimension $d=2$.
Here we generalize these results to arbitrary $d$ using MF theory
and a mapping onto a Ginzburg-Landau model.

The reason for the appearance of normal critical behavior
is a subtle interplay between the symmetry with respect
to sublattice ordering and broken translational invariance
due to the free surface.
Consider first a finite system with {\em periodic\/} boundary conditions.
The precise form of the Hamiltonian ${\cal H}\{\sigma_{\bbox{i}}\}$
does not matter here and will be given in Sec.\ \ref{MFTasNLM}.
The spin variable $\sigma_{\bbox{i}}=1$ ($\sigma_{\bbox{i}}=-1$)
represents an $A$ ($B$) atom on lattice site $\bbox{i}$.
The statistical weight of a configuration $\{\sigma_{\bbox{i}}\}$
is given by the finite-volume Gibbs measure
\BE\label{Gibbs}
\rho(\{\sigma_{\bbox{i}}\}) =
\frac{1}{Z}e^{-\beta{\cal H}\{\sigma_{\bbox{i}}\}}\,,\quad
\beta\equiv1/(k_BT)\,,
\EE
where $T$ is the temperature and $k_B$ denotes Boltzmann's constant.
The normalization factor $Z$ is the grand-canonical partition function.
The local concentration $c_{\bbox{i}}$ of $A$ atoms at site $\bbox{i}$
can be expressed in terms of the mean magnetization
$\langle\sigma_{\bbox{i}}\rangle$
as $c_{\bbox{i}}=(1+\langle\sigma_{\bbox{i}}\rangle)/2$.
The Gibbs measure (\ref{Gibbs}) is translationally invariant,
\BE
\rho(\{\sigma_{\bbox{i}}\})=\rho(\{\sigma_{\bbox{i}}^\prime\})\,,\quad
\sigma_{\bbox{i}}^\prime\equiv\sigma_{\bbox{i}+\bbox{t}}\,,
\EE
where $\bbox{t}$ may be chosen arbitrarily from the set $\cal T$
of all bcc lattice vectors. Due to {\em spontaneous symmetry breaking},
the thermodynamic states that are obtained by calculating
expectation values with the measure (\ref{Gibbs}) and taking
the infinite-volume limit need {\em not\/} be translationally invariant.
(Formally, one introduces a symmetry-breaking ``staggered'' field which
is sent to zero {\em after\/} the thermodynamic limit has been performed.)
In the ordered phase below $T_c$, one has
$\langle\sigma_{\bbox{i}}\rangle\neq\langle\sigma_{\bbox{i}+\bbox{t}}\rangle$
for all translations $\bbox{t}=\bbox{t}_{\alpha\beta}$
that map $\alpha$ to $\beta$-sites (Fig.\ \ref{Fig_B2}).

Surfaces are introduced by imposing {\em free\/} boundary conditions along
one direction while retaining periodic boundary conditions in the other
directions. Then the measure (\ref{Gibbs}) is invariant under the subset
${\cal T}^\prime\subset{\cal T}$ of translations {\em parallel\/}
to the surface.
We call the surface {\em symmetry-preserving\/} if ${\cal T}^\prime$
contains a ``sublattice-exchanging'' translation $\bbox{t}_{\alpha\beta}$,
and {\em symmetry-breaking\/} otherwise. In the case of the bcc lattice
considered here, symmetry-breaking surfaces are characterized
by the alternation of $\alpha$ and $\beta$-planes along the direction normal
to the surface. Let us assume that no spontaneous symmetry breaking takes
place above $T_c$, which would require supercritically enhanced surface
couplings, so that
$\langle\sigma_{\bbox{i}}\rangle=\langle\sigma_{\bbox{i}+\bbox{t}}\rangle$
for $T\ge T_c$ and $\bbox{t}\in{\cal T}^\prime$.

Thus for {\em symmetry-preserving\/} surfaces and $T\ge T_c$,
both sublattice magnetizations are the same within each plane parallel
to the surface, and the {\em OP profile vanishes}.
Nonetheless one obtains an inhomogeneous magnetization (or concentration)
profile due to surface segregation. E.~g., segregation of one component
at the symmetry-preserving (110) surface leads to an alternation
of $A$ and $B$-rich lattice planes since the interactions favor
the occupation of nearest-neighbor (NN) sites by different species.
As will be shown in Sec.\ \ref{LinearMF}, the scale on which the segregation
profile decays to its bulk value is on the order of the lattice
constant even near $T_c$. This length should be irrelevant
(in the renormalization-group sense), so that the surface
must exhibit {\em ordinary\/} critical behavior.

{\em Symmetry-breaking\/} surfaces, like the (100) surface,
destroy the two-sublattice symmetry.
Surface segregation again leads to an inhomogeneous
concentration profile for $T\ge T_c$, which is now equivalent
to a {\em nonvanishing OP profile\/} since adjacent lattice planes belong
to different sublattices. The OP profile decays on the scale
of the bulk correlation length, which {\em diverges\/} for $T\to T_c$
(Sec.\ \ref{LinearMF}). Such a persistence of surface order for $T\ge T_c$
has been confirmed in recent experiments on FeCo(100).\cite{FeCo}

According to the experimental results of Ref.\ \onlinecite{FeCo},
supercritical enhancement of the surface couplings can be ruled out.
Thus it is natural to attribute the persistence of surface order
to an {\em ordering surface field\/} $g_1\neq0$. This field causes the
system to display the {\em normal\/} transition. However,
for the binary alloys considered here an ordering field corresponds
to a local chemical potential acting differently on the two sublattices
(staggered field in magnetic language).
There is no natural source for such a field on the microscopic level.
The challenge is to demonstrate in an unequivocal fashion that a nonzero $g_1$
nevertheless emerges in a continuum (coarse-grained) description,
i.~e., in the context of a Ginzburg-Landau model, and to derive
a MF expression for $g_1$ in terms of the lattice model parameters. Of course,
in comparing theory and experimental or simulation data,
one should keep in mind that $g_1$ may be small, so that the crossover
to normal critical behavior occurs only close to $T_c$.\cite{MC}

In the next section, we shall reformulate the lattice MF equations
for the semi-infinite alloy with free (100) and (110) surfaces
as a problem in discrete dynamics, i.~e., the iteration of nonlinear
symplectic maps.\cite{PW} From the linearized maps the characteristic length
scales of both the concentration and OP profiles will be calculated
(Sec.\ \ref{LinearMF}). The full nonlinear maps will be analyzed
in Sec.\ \ref{latticeMFT_num}. After the derivation
of the Ginzburg-Landau model for the (100) surface
in Sec.\ \ref{GL}, we shall compare the predictions of the continuum theory
with the numerical solutions of the lattice MF equations (Sec.\ \ref{normal}).
This will serve to demonstrate that the (100) surface displays
{\em normal\/} critical behavior generically. However,
for certain exceptional parameter values, where $g_1$ happens to vanish
at $T=T_c$, one recovers the singularities of the ordinary transition
(Sec.\ \ref{ordinary}). We will summarize our main results
in Sec.\ \ref{Summary}.


\section{MF equations as nonlinear maps}\label{MFTasNLM}
We consider the lattice-gas model of a binary ($AB$) alloy on a bcc lattice.
Each atomic configuration is characterized by the values
of the occupation variables $p_{\bbox{i}}^A$, $p_{\bbox{i}}^B$,
where $p_{\bbox{i}}^\nu=1$
if site $\bbox{i}$ is occupied by an atom of type $\nu\in\{A,B\}$
and $p_{\bbox{i}}^\nu=0$ otherwise.
Within the grand-canonical ensemble, the configurational energy reads
\BE\label{lattice_gas}
E\left\{p_{\bbox{i}}^A,p_{\bbox{i}}^B\right\}
= \frac{1}{2}\sum_{\bbox{i}\neq\bbox{j}}\sum_{\nu,\tau}
\epsilon_{\bbox{i}\bbox{j}}^{\nu\tau}p_{\bbox{i}}^\nu p_{\bbox{j}}^\tau
- \sum_{\nu}\mu_\nu\sum_{\bbox{i}}p_{\bbox{i}}^\nu\,,
\EE
where
$\epsilon_{\bbox{i}\bbox{j}}^{\nu\tau}=\epsilon_{\bbox{i}\bbox{j}}^{\tau\nu}$
is the interaction energy between $\nu$ and $\tau$ atoms at sites
$\bbox{i}$ and $\bbox{j}$, and $\mu_A$ and $\mu_B$ are chemical potentials
for $A$ and $B$ atoms, respectively. We neglect vacancies, so that
$p_{\bbox{i}}^A+p_{\bbox{i}}^B=1$, and rewrite the occupation variables
in terms of Ising spins $\sigma_{\bbox{i}}=\pm1$ as
\[
p_{\bbox{i}}^A = \frac{1}{2}\left(1+\sigma_{\bbox{i}}\right)\,,\quad
p_{\bbox{i}}^B = \frac{1}{2}\left(1-\sigma_{\bbox{i}}\right)\,.
\]
Then (\ref{lattice_gas}) takes the form of an Ising Hamiltonian
\BE\label{Hamiltonian}
{\cal H}\{\sigma_{\bbox{i}}\} = -\frac{1}{2}\sum_{\bbox{i}\neq\bbox{j}}J_{\bbox{i}\bbox{j}}
\sigma_{\bbox{i}}\sigma_{\bbox{j}}
- \sum_{\bbox{i}}H_{\bbox{i}}\sigma_{\bbox{i}}\,,
\EE
where a spin-independent term has been dropped and
\BA
J_{\bbox{i}\bbox{j}} &=& \frac{1}{4}\left(2\epsilon_{\bbox{i}\bbox{j}}^{AB}
-\epsilon_{\bbox{i}\bbox{j}}^{AA}-\epsilon_{\bbox{i}\bbox{j}}^{BB}\right)\,,
\eqnum{\ref{Hamiltonian}a}\\
H_{\bbox{i}} &=& \frac{1}{4}\sum_{\bbox{j}\,(\neq\bbox{i})}
\left(\epsilon_{\bbox{i}\bbox{j}}^{BB}-\epsilon_{\bbox{i}\bbox{j}}^{AA}\right)
+ \frac{1}{2}\left(\mu_A-\mu_B\right)\,.\eqnum{\ref{Hamiltonian}b}
\EA
In the following we only consider NN interactions
$\epsilon^{AA}$, $\epsilon^{AB}$, and $\epsilon^{BB}$.
Moreover, we do not allow for enhanced surface interactions.
For an $B2$-$A2$ order-disorder transition to exist, the Ising coupling
$J=(2\epsilon^{AB}-\epsilon^{AA}-\epsilon^{BB})/4$ must be
{\em antiferromagnetic\/} ($J<0$).
For semi-infinite systems with (100) or (110) surfaces the local field
(\ref{Hamiltonian}b) differs from its bulk value only in the first layer,
\BE\label{H,H1}
H_{\bbox{i}} = \left\{\begin{array}{l@{\quad}l}
H + H_1 & \text{if $\bbox{i}\in$ surface}\,,\\
H       & \text{otherwise}\,,
\end{array}\right.
\EE
where
\BA
H &=& \frac{\zeta}{4}\left(\epsilon^{BB}-\epsilon^{AA}\right)
+\frac{1}{2}\left(\mu_A-\mu_B\right)\,,\eqnum{\ref{H,H1}a}\\
H_1 &=& \frac{\zeta_s-\zeta}{4}\left(\epsilon^{BB}-\epsilon^{AA}\right)\,.
\eqnum{\ref{H,H1}b}
\EA
Here, $\zeta$ and $\zeta_s$ are the coordination numbers
of bulk and surface spins, i.~e., $\zeta=8$, while $\zeta_s=4$ and $\zeta_s=6$
for the (100) and (110) surface, respectively.

Some remarks about the role of the surface field $H_1$ are in order here.
The field $H_1$ favors one component at the surface and thus accounts
for surface segregation (see Sec.\ \ref{Intro}). Because of different
interaction energies $\epsilon^{AA}\neq\epsilon^{BB}$ it is nonzero
generically. More generally, $H_1$ also models other effects
such as different sizes of the two constituents.
For symmetry-preserving orientations,
$H_1$ acts uniformly on $\alpha$ and $\beta$-sites
at the surface and must not be confused with an ordering (staggered) field.
For symmetry-breaking surfaces, spins on $\alpha$ and $\beta$-sites
in the first two layers experience {\em different\/} fields $H+H_1$ and $H$.
Hence $H_1$ should contribute to an ``effective'' staggered surface field
$g_1\neq0$. However, even if $H_1=0$ (but $H\neq0$) one obtains
an inhomogeneous (oscillating) magnetization profile equivalent
to a nonzero local order parameter, and an ordering surface field
$g_1\neq0$ should again emerge in a coarse-grained description.

The mean-field (MF) or Bragg-Williams approximation is conveniently
formulated in terms of a variational principle.\cite{Ducastelle}
The free-energy functional reads
\BE\label{MF_fctl}
F_{\text{MFA}}\{\langle\sigma_{\bbox{i}}\rangle\} =
{\cal H}\{\langle\sigma_{\bbox{i}}\rangle\}
- TS_{\text{MFA}}\{\langle\sigma_{\bbox{i}}\rangle\}\,,
\EE
with the MF entropy
\[
S_{\text{MFA}}\{\langle\sigma_{\bbox{i}}\rangle\} = -k_B\sum_{\bbox{i}}
\int_0^{\langle\sigma_{\bbox{i}}\rangle}dx\,\tanh^{-1}x\,.
\]
Variation of $F_{\text{MFA}}\{\langle\sigma_{\bbox{i}}\rangle\}$ yields
the MF equations
\BE\label{MFeqs}
\langle\sigma_{\bbox{i}}\rangle =
\tanh\!\left[\frac{1}{k_BT}\left(H_{\bbox{i}}+\sum_{\bbox{j}\neq\bbox{i}}
J_{\bbox{i}\bbox{j}}\,\langle\sigma_{\bbox{j}}\rangle\right)\right]\,.
\EE
The magnetization densities of the two sublattices vary only in the direction
perpendicular to the surface. For the (100) surface one may thus write
\[
\langle\sigma_{\bbox{i}}\rangle = m_n \quad
\text{for $\bbox{i}\in$ lattice plane $n$}\,,
\]
where $m_n$ is the magnetization density of lattice plane $n$.
Likewise one has, for the (110) orientation,
\[
\langle\sigma_{\bbox{i}}\rangle =
\left\{\begin{array}{l@{\quad}l}
m_{n,\alpha} & \text{for $\bbox{i}\in$ plane $n$, sublattice $\alpha$}\,,\\
m_{n,\beta} & \text{for $\bbox{i}\in$ plane $n$, sublattice $\beta$}\,.
\end{array}\right.
\]
It is convenient to introduce the reduced quantities
\[
K\equiv\frac{4|J|}{k_BT}\,,\quad h\equiv{H\over4|J|}\,,\quad
h_1\equiv{H_1\over4|J|}\,.
\]
Then the MF equations read, for the (100) surface,
\BE\label{MF100}
\frac{1}{K}\tanh^{-1}m_n = h - m_{n-1} - m_{n+1}
\EE
for $n>1$, and
\BE\label{MF100_surf}
\frac{1}{K}\tanh^{-1}m_1 = h + h_1 - m_2\,.
\EE
For the (110) orientation, one has
\BA\label{MF110}\addtocounter{equation}{1}
\frac{1}{K}\tanh^{-1}m_{n,\alpha} &=& h - m_{n,\beta}
 - \frac{m_{n-1,\beta}+m_{n+1,\beta}}{2}\,,\eqnum{\theequation a}\\
\frac{1}{K}\tanh^{-1}m_{n,\beta} &=& h - m_{n,\alpha}
 - \frac{m_{n-1,\alpha}+m_{n+1,\alpha}}{2}\eqnum{\theequation b}
\EA
for $n>1$, and
\BA\label{MF110_surf}\addtocounter{equation}{1}
\frac{1}{K}\tanh^{-1}m_{1,\alpha} &=&
h + h_1 - m_{1,\beta} - \frac{1}{2}m_{2,\beta}\,,\eqnum{\theequation a}\\
\frac{1}{K}\tanh^{-1}m_{1,\beta} &=&
h + h_1 - m_{1,\alpha} - \frac{1}{2}m_{2,\alpha}\,.\eqnum{\theequation a}
\EA
We now combine the magnetization densities of two neighboring planes
into single points in ${\Bbb R}^2$ and ${\Bbb R}^4$, respectively,
\BAA
\bbox{v}_n &\equiv& (m_{n-1},m_n)^T\,,\\
\bbox{w}_n &\equiv&
(m_{n-1,\alpha},m_{n-1,\beta},m_{n,\alpha},m_{n,\beta})^T\,,
\EAA
where ``$T$'' denotes the transpose.
Then (\ref{MF100}) and (\ref{MF110}) are equivalent to the recursion equations
\BA
\label{DynSys100}
\bbox{v}_{n+1} &=& F(\bbox{v}_n)\,,\\
\label{DynSys110}
\bbox{w}_{n+1} &=& G(\bbox{w}_n)\,,
\EA
where the nonlinear maps $F$ and $G$ are defined by
\[
F:\;\left(\begin{array}{l}x\\y\end{array}\right)
\rightarrow
\left(\begin{array}{l}
y\\ h - x - \frac{1}{K}\tanh^{-1}y \end{array}
\right)\,,
\]
and
\[
G:\;\left(\begin{array}{l}x_1\\ x_2\\ x_3\\ x_4\end{array}\right)\rightarrow
\left(\begin{array}{l}x_3\\ x_4\\
2h - x_1 - 2x_3 - \frac{2}{K}\tanh^{-1}x_4\\
2h - x_2 - 2x_4 - \frac{2}{K}\tanh^{-1}x_3\end{array}\right)\,.
\]
One advantage of rewriting the MF equations in terms of the discrete dynamics
(\ref{DynSys100}) and (\ref{DynSys110}) is that one may gain an overview
of {\em all} solutions by iterating arbitrary starting points $\bbox{v}_1$
and $\bbox{w}_1$. In this way one obtains trajectories
$\bbox{v}_1,\bbox{v}_2,\ldots$ and $\bbox{w}_1,\bbox{w}_2,\ldots$
in a two-dimensional (2D) and four-dimensional (4D) phase space, respectively.
The maps $F$ and $G$ are both {\em symplectic}, i.~e., their differentials
$DF$ and $DG$ are symplectic matrices, and thus generate a discrete
{\em Hamiltonian\/} dynamics on these phase spaces.
Note that any symplectic map is volume-preserving, in particular.
The theory of nonlinear dynamics offers convenient tools to understand
the discrete dynamics generated by such maps.\cite{Arrowsmith}

The MF equations (\ref{MF100_surf}) and (\ref{MF110_surf})
take the same form as (\ref{MF100}) and (\ref{MF110}) if fictitious
zeroth layer magnetizations $m_0=-h_1$ and $m_{0,\alpha}=m_{0,\beta}=-2h_1$
are introduced, i.~e.,
\BA
\bbox{v_2} &=& F(\bbox{v}_1)\,,\\
\bbox{w_2} &=& G(\bbox{w}_1)\,,
\EA
where $\bbox{v}_1$ and $\bbox{w}_1$ satisfy the boundary conditions
\BA
\label{BC_surf100}
\bbox{v}_1 &=& (-h_1,m_1)^T\,,\\
\label{BC_surf110}
\bbox{w}_1 &=& (-2h_1,-2h_1,m_{1,\alpha},m_{1,\beta})^T\,.
\EA
Moreover we require that the sublattice magnetization densities approach
their bulk values for $n\to\infty$,
\BE\label{BC_infty}
\begin{array}{rcl}
m_{2n-1} \to m_\alpha &,& m_{2n}\to m_\beta\,,\\
m_{n,\alpha} \to m_\alpha &,& m_{n,\beta}\to m_\beta\,,
\end{array}
\EE
where $m_\alpha,m_\beta$ are the solutions of the bulk MF equations
[Eqs.\ (\ref{MFbulk})]. As will be discussed in the next section,
the bulk solutions correspond
to fixed points of the maps $F$ and $G$. Then (\ref{BC_infty}) implies
that the trajectories converge to these fixed points.
The task of solving the MF equations for the semi-infinite system
(for given values of $K$, $h$, and $h_1$)
thus translates into finding the intersections of the {\em stable manifold\/}
(inset) of the corresponding fixed point with the surface boundary conditions
(\ref{BC_surf100}) and (\ref{BC_surf110}) (see Sec.\ \ref{latticeMFT_num}).

We finally quote an important {\em symmetry property\/} of the above maps.
The MF equations (\ref{MF100}) and (\ref{MF110})
are symmetric with respect to interchanging the layer magnetizations
of the planes $n-1$ and $n+1$. Thus one has
\[
F\left(\begin{array}{c}m_{n+1}\\m_n\end{array}\right) =
\left(\begin{array}{c}m_{n}\\m_{n-1}\end{array}\right)\,,
\]
and an analogous relation for $G$. It follows that both maps are invertible
and that their inverses $F^{-1}$ and $G^{-1}$ are given by
\BE\label{refl_symm}
\begin{array}{r@{\;=\;}l}
\D F^{-1} & \D R \circ F \circ R\,,\\
\D G^{-1} & \D S \circ G \circ S\,,
\end{array}
\EE
where
\BAA
R\,:\, (x,y)^T &\rightarrow& (y,x)^T\,,\\
S\,:\, (x_1,x_2,x_3,x_4)^T &\rightarrow& (x_3,x_4,x_1,x_2)^T\,.
\EAA


\section{Analysis of Nonlinear Maps}\label{NLmaps}
\subsection{Linearized MF equations and length scales}\label{LinearMF}
The bulk MF equations (\ref{MFbulk}) are equivalent
to the fixed point equations of the nonlinear maps $G$ and $F^2=F\circ F$,
the second iterate of $F$. For $T\neq T_c$, linearization of the maps
about the fixed points yields the asymptotic (exponential) decay
of the sublattice magnetization profiles away from the surface.
The decay lengths can be expressed by the eigenvalues of the linearized maps.
In the case of the (100) surface only {\em one\/} length scale,
proportional to the OP correlation length, governs the decay
of both the sublattice magnetization and the OP profiles.
An additional length associated with the mean magnetization density
occurs for the (110) orientation.
However, this length scale stays {\em finite\/} at $T=T_c$.

\begin{figure}[b]
\epsffile{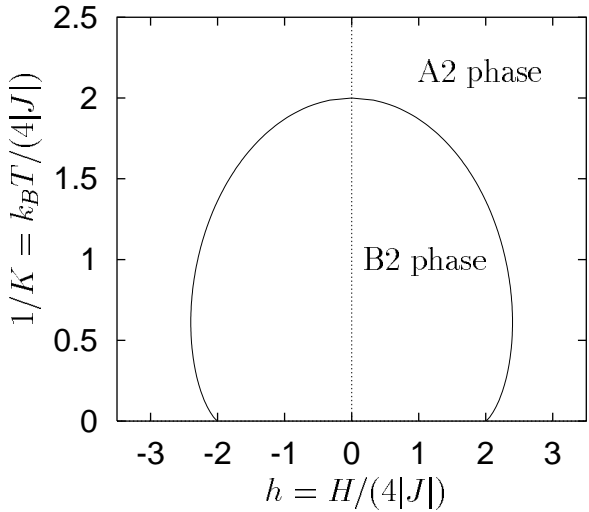}
\vspace{2mm}
\caption
{
MF phase diagram of the NN Ising antiferromagnet on a bcc lattice,
showing a line of continuous transitions between the disordered
(A2) and the ordered (B2) phase.
}
\label{Fig_phdiagr}
\end{figure}

An analysis of the bulk MF equations is straightforward and may be found
in Appendix \ref{App_bulk}. The critical coupling $K_c=(4|J|)/(k_BT_c)$
as a function of the uniform bulk field $h$ is determined by
\BE\label{Kc}
\frac{1}{2K_c(h)}=1-m_c(h)^2\,,
\EE
where $m_c=m_c(h)$ is the uniform magnetization at $T=T_c$. Using (\ref{mdis})
to eliminate $m_c$ in favor of $K_c$ and $h$ in (\ref{Kc}), one obtains
an expression for the critical line (Fig.\ \ref{Fig_phdiagr}).
\cite{bulkphdiagr,DB}

\begin{figure}[t]
\centerline{\epsffile{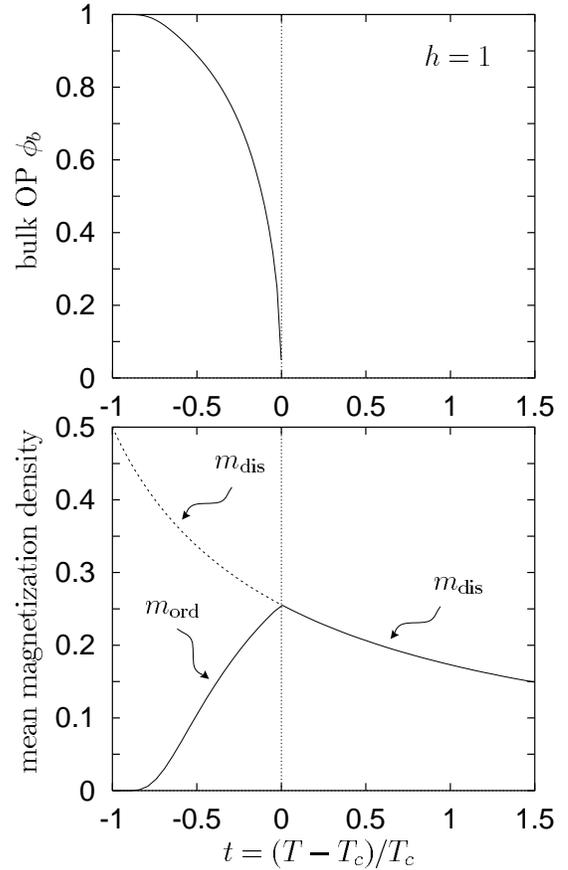}}
\vspace{2mm}
\caption
{
Temperature dependence of the bulk OP and mean magnetization density.
Below $T_c$, the disordered state (dashed line)
becomes thermodynamically unstable.
}
\label{Fig_OPmbulk}
\end{figure}

The bulk sublattice magnetization densities may be written
\BE\label{mphi}
m_\alpha=m+\phi\,,\quad m_\beta=m-\phi\,,
\EE
where $m$ and $\phi$ are the mean magnetization and the OP, respectively.
In the region of the phase diagram where the A2 phase is thermodynamically
stable, the bulk MF equations have a unique solution (see Appendix
\ref{App_bulk})
\BE\label{dis}
m=m_{\text{dis}}(K,h)\,,\quad\phi=0\,.
\EE
Thus the only fixed points of $F^2$ and $G$ in this case are
\BE\label{FP_dis}
\begin{array}{r@{\;=\;}l}
\D\bbox{v}_{\text{dis}} & \D(m_{\text{dis}},m_{\text{dis}})^T\,,\\
\D\bbox{w}_{\text{dis}} &
\D(m_{\text{dis}},m_{\text{dis}},m_{\text{dis}},m_{\text{dis}})^T\,.
\end{array}
\EE
The solution (\ref{dis}) becomes thermodynamically unstable on crossing
the critical line (Fig.\ \ref{Fig_phdiagr}). At the same time,
two new solutions describing the pure B2 phases emerge,
\BE\label{ord}
m=m_{\text{ord}}(K,h)\,,\quad\phi=\pm\phi_b(K,h)\neq0\,,
\EE
corresponding to the fixed points
\BE\label{FP_ord}
\begin{array}{r@{\;=\;}l}
\D\bbox{v}_{\text{ord}}^{1,2} &
\D(m_{\text{ord}}\mp\phi_b,m_{\text{ord}}\pm\phi_b)^T\,,\\
\D\bbox{w}_{\text{ord}}^{1,2} &
\D(m_{\text{ord}}\pm\phi_b,m_{\text{ord}}\mp\phi_b,
m_{\text{ord}}\pm\phi_b,m_{\text{ord}}\mp\phi_b)^T\,.
\end{array}
\EE
Note that the fixed points $\bbox{v}_{\text{ord}}^1$, $\bbox{v}_{\text{ord}}^2$
form a 2-cycle of the map $F$,
\BE
F\left(\bbox{v}_{\text{ord}}^1\right) = \bbox{v}_{\text{ord}}^2\,,\quad
F\left(\bbox{v}_{\text{ord}}^2\right) = \bbox{v}_{\text{ord}}^1\,.
\EE
The bulk OP and the mean magnetization density at fixed bulk field $h=1$
are shown in Fig.\ \ref{Fig_OPmbulk} as a function of the reduced temperature
\[
t\equiv\frac{T-T_c}{T_c}=\frac{K_c-K}{K}\,.
\]
The OP vanishes as $|t|^{\beta}$ for $t\to0-$, with the usual MF exponent
$\beta=1/2$ [Eq.\ (\ref{mphi_exp})]. The mean magnetization density
is a nonordering (or noncritical) density. Typically, such quantities
display $|t|^{1-\alpha}$ singularities, where $\alpha$ is the bulk specific
heat exponent. In MF theory this behavior reduces to a discontinuity
in the first temperature derivative since $\alpha=0$
[see Eqs.\ (\ref{mdis_exp}) and (\ref{mphi_exp})].

The linearization of $F^2$ about any of the fixed points
$\bbox{v}_{\text{dis}}$ and $\bbox{v}_{\text{ord}}^{1,2}$ has the eigenvalues
\BE\label{evals100}
\lambda_{1,2}=-1+2u_\alpha u_\beta
\pm2\sqrt{u_\alpha u_\beta}\sqrt{u_\alpha u_\beta-1}
\EE
and the associated eigenvectors
\BE\label{evecs100}
\bbox{l}_{1,2} = \left(1,\,-\frac{1+\lambda_{1,2}}{2u_\beta}\right)^T\,,
\EE
where $u_\alpha$ and $u_\beta$ are defined by (\ref{uab}). Since $F$
is symplectic and thus area-preserving (Sec.\ \ref{MFTasNLM}), one has
$\lambda_1\lambda_2=1$. Likewise, the eigenvalues of the linearization
of $G$ about $\bbox{w}_{\text{dis}}$ or $\bbox{w}_{\text{ord}}^{1,2}$ are
\BE\label{evals110}
\begin{array}{r@{\;=\;}l}
\D\Lambda_{1,2} & \D -1 + 2\sqrt{u_\alpha u_\beta}
\pm2\sqrt{u_\alpha u_\beta-\sqrt{u_\alpha u_\beta}}\,,\\
\D\Lambda_{3,4} & \D -1 - 2\sqrt{u_\alpha u_\beta}
\pm2\sqrt{u_\alpha u_\beta+\sqrt{u_\alpha u_\beta}}\,,\end{array}
\EE
with the corresponding eigenvectors
\BE
\begin{array}{r@{\;=\;}r@{,\,}r@{,\,}r@{,\,}r}
\D\bbox{L}_{1,2} & \D\bigg(1 & \D-\sqrt{\frac{u_\alpha}{u_\beta}} &
\D\Lambda_{1,2} & \D-\sqrt{\frac{u_\alpha}{u_\beta}}\Lambda_{1,2}\bigg)^T\,,\\
\D\bbox{L}_{3,4} & \D\bigg(1 & \D\sqrt{\frac{u_\alpha}{u_\beta}} &
\D\Lambda_{3,4} & \D\sqrt{\frac{u_\alpha}{u_\beta}}\Lambda_{3,4}\bigg)^T\,.
\end{array}
\EE
Since $G$ is symplectic, the eigenvalues come in pairs
($\Lambda_1$, $\Lambda_2$) and ($\Lambda_3$, $\Lambda_4$)
with $\Lambda_1\Lambda_2=\Lambda_3\Lambda_4=1$.
Repeated application of the linearized map to $\bbox{L}_{1,2}$ generates
eigensolutions with opposite sublattice magnetization densities
within each layer. Therefore,
$\bbox{L}_{1,2}$ represent ``ordering'' eigenmodes. The eigensolutions
generated by $\bbox{L}_{3,4}$ show an oscillating profile of the magnetization
density because $\Lambda_{3,4}<0$. However, the local magnetization
is the same for {\em all\/} sites in a given lattice plane parallel
to the surface. Hence the OP profile vanishes and one may refer
to $\bbox{L}_{3,4}$ as ``nonordering'' eigenmodes.

\begin{figure}
\centerline{\epsffile{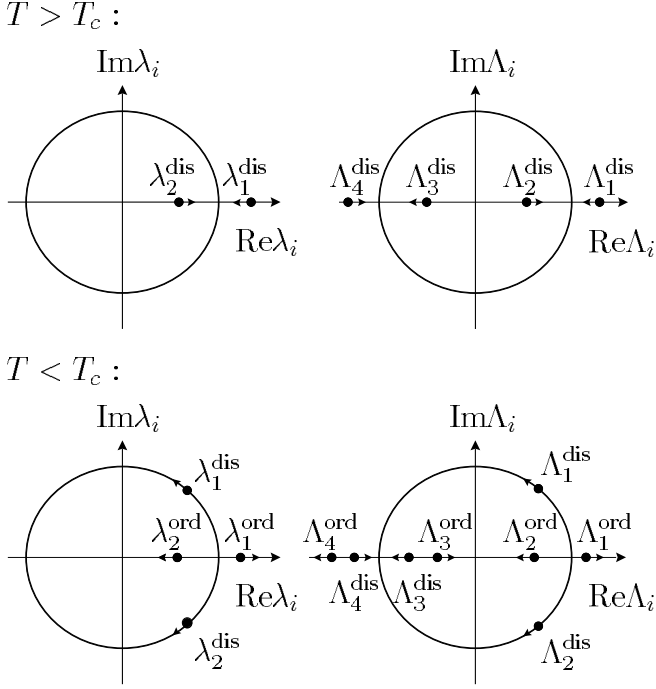}}
\vspace{2mm}
\caption
{
Behavior of the eigenvalues (\protect\ref{evals100})
and (\protect\ref{evals110}) in the complex plane as one crosses
the critical line (cf.\ Fig.\ \protect\ref{Fig_phdiagr}).
The superscripts ``dis'' and ``ord''
indicate the linearizations about the fixed points (\protect\ref{FP_dis})
and (\protect\ref{FP_ord}), respectively.
}
\label{Fig_evals}
\end{figure}
The behavior of the eigenvalues in the complex plane
as one crosses the critical line is shown schematically
in Fig.\ \ref{Fig_evals}. The eigenvalues $\lambda_1$, $\lambda_2$
collide at $+1$ and form a complex conjugate pair on the unit circle
in the ordered phase. Thus the character of the fixed point
$\bbox{v}_{\text{dis}}$ changes from hyperbolic to elliptic. At the same time,
two new real eigenvalues corresponding to the hyperbolic fixed points
$\bbox{v}_{\text{ord}}^{1,2}$ emerge. Within the notions
of nonlinear dynamics, the map $F$ undergoes a period-doubling bifurcation.
The eigenvalues $\Lambda_1$, $\Lambda_2$ of the 4D map $G$ show
an analogous behavior. However, the fixed point $\bbox{w}_{\text{dis}}$
remains unstable in the ordered phase since $\Lambda_3$, $\Lambda_4$ stay real.

The solutions of the linearized MF equations
satisfying the bulk boundary conditions (\ref{BC_infty}) read
\BAA
\bbox{v}_{2n+1} &=& \bbox{v}_\ast + a_2\lambda_2^n\bbox{l}_2\,,\\
\bbox{w}_{n+1} &=& \bbox{w}_\ast + A_2\Lambda_2^n\bbox{L}_2
+ A_3\Lambda_3^n\bbox{L}_3\,,
\EAA
where $\bbox{v}_\ast$ and $\bbox{w}_\ast$ stand for one of the fixed points
(\ref{FP_dis}) and (\ref{FP_ord}).
The coefficients $a_2$, $A_2$, and $A_3$ are fixed by the surface boundary
conditions (\ref{BC_surf100}) and (\ref{BC_surf110}).

Thus for the (100) orientation we obtain
\BE\label{profiles100}
\begin{array}{r@{\;=\;}l}
\D m_{2n+1} & \D m_\alpha+a\text{e}^{-(2n+1)/\xi}\,,\\
\D m_{2n} & \D m_\beta-b\text{e}^{-2n/\xi}\,,
\end{array}
\EE
with the decay length
\BE
\xi = 2\left|\ln\lambda_2\right|^{-1}\,,\eqnum{\ref{profiles100}a}
\EE
and the amplitudes
\BE
a = (h_1+m_\beta)\sqrt{\frac{u_\beta}{u_\alpha}}\,,\quad
b = h_1+m_\beta\,.\eqnum{\ref{profiles100}b}
\EE
In the disordered phase, the amplitudes simplify to
\BE
a = b = h_1 + m_{\text{dis}}\,.\eqnum{\ref{profiles100}b$^\prime$}
\EE
We define the local OP $\phi_n$ by
\BE\label{OPasym}
\phi_n\equiv{1\over2}(-1)^n\left(m_{n+1}-m_n\right)\,,
\EE
where the power of $-1$ ensures that one always subtracts the magnetization
densities of $\beta$-planes from those of $\alpha$-planes.\cite{def-OP}
Eqs.\ (\ref{profiles100}) imply a nonvanishing OP profile, which decays
on the scale of $\xi$. In fact, $\xi$ may be identified,
up to a proportionality factor, with the bulk OP correlation length.
To see this, one expands $u_\alpha u_\beta$ in Eq.\ (\ref{evals100})
in powers of $t$ using (\ref{mdis_exp}) and (\ref{mphi_exp}). This gives
\BE\label{xi_exp}
\xi=\xi_{\pm}|t|^{-1/2} + {\cal O}(t)\,,
\EE
where
\BA
\xi_+ &=& \xi_+(h) = \frac{1}{2\sqrt{2-2K_c(h)m_c(h)[h-2m_c(h)]}}\,,
\eqnum{\ref{xi_exp}a}\\
\xi_- &=& \xi_-(h) = \frac{1}{\sqrt{2}}\xi_+(h)\,.\eqnum{\ref{xi_exp}b}
\EA
The decay length displays a $|t|^{-\nu}$ singularity just
as the bulk correlation length, with the MF exponent $\nu=1/2$.
Asymptotically, $\xi$ should thus be proportional to the correlation length.
Indeed, one has $\xi_+(h)/\xi_-(h)=\sqrt{2}$, which is the MF value
of the universal amplitude ratio of the correlation lengths above and below
$T_c$.\cite{ampl-ratio} As $t\to0$, the exponential decay of the OP profile
becomes a power law, whose precise form will be investigated in Sec.\ \ref{GL}.

Likewise, the magnetization profiles for the (110) orientation are
\BE\label{profiles110}
\begin{array}{r@{\;=\;}l}
\D m_{n,\alpha} & \D m_\alpha + A\text{e}^{-n/\xi^\prime}
+ (-1)^n\tilde{A}\text{e}^{-n/\tilde{\xi}}\,,\\
\D m_{n,\beta} & \D m_\beta - B\text{e}^{-n/\xi^\prime}
+ (-1)^n\tilde{B}\text{e}^{-n/\tilde{\xi}},
\end{array}
\EE
where now {\em two\/} length scales appear,
\BE
\xi^\prime=\left|\ln\Lambda_2\right|^{-1}\,,\quad
\tilde{\xi}=\left|\ln\left|\Lambda_3\right|\right|^{-1}\,.
\eqnum{\ref{profiles110}a}
\EE
The amplitudes are given by
\BE
\begin{array}{r@{\;=\;}l}
\D A & \D-\left(1-\sqrt{\frac{u_\beta}{u_\alpha}}\right)h_1
-\frac{1}{2}\left(m_\alpha-\sqrt{\frac{u_\beta}{u_\alpha}}m_\beta\right)\,,\\
\D\tilde{A} & \D-\left(1+\sqrt{\frac{u_\beta}{u_\alpha}}\right)h_1
-\frac{1}{2}\left(m_\alpha+\sqrt{\frac{u_\beta}{u_\alpha}}m_\beta\right)\,,\\
\D B & \D\sqrt{\frac{u_\alpha}{u_\beta}}A\,,\quad
\tilde{B} = \sqrt{\frac{u_\alpha}{u_\beta}}\tilde{A}\,.
\end{array}\eqnum{\ref{profiles110}b}
\EE
In the disordered phase above $T_c$, they simplify to
\BE
A = B = 0\,,\quad \tilde{A} = \tilde{B} = -2h_1-m_{\text{dis}}
\eqnum{\ref{profiles110}b$^\prime$}\,.
\EE
In particular, the OP profile
\BE
\phi_n\equiv\frac{1}{2}(m_{n,\alpha}-m_{n,\beta})
\EE
vanishes for $T\ge T_c$, which is a consequence of the symmetry of the (110)
surface with respect to the two sublattices and the fact that neither
enhanced surface couplings nor a staggered surface field are present.
Asymptotically, $\xi^\prime$ and $\tilde{\xi}$ behave as
\BA\addtocounter{equation}{1}
\xi^\prime &=& \frac{1}{\sqrt{2}}\xi_\pm|t|^{-1/2} + {\cal O}(t)\,,
\eqnum{\theequation a}\\
\tilde{\xi} &=& \left|\ln\left(3-2\sqrt{2}\right)\right|^{-1} + {\cal O}(t)\,,
\eqnum{\theequation b}\label{length_scale110}
\EA
with $\xi_\pm$ given in (\ref{xi_exp}) and
$\left|\ln\left(3-2\sqrt{2}\right)\right|^{-1}\simeq0.57$.
The length $\xi^\prime$ associated with the ordering eigenmodes
diverges as $t\to0$ and may again be identified with the bulk correlation
length (up to a proportionality constant),\cite{xi} whereas $\tilde{\xi}$
stays on the order of the lattice constant. As will be seen
in the next section, $\tilde{\xi}$ describes the decay
of the mean magnetization profile for $T\ge T_c$.

Note that for both surface orientations the layer magnetizations
oscillate about $m_{\text{dis}}$ for $T>T_c$
due to the antiferromagnetic coupling between adjacent lattice planes.
However, only in the case of the (100) orientation does
this oscillating profile lead to a nonvanishing OP profile
whose characteristic length scale diverges as $t\to0+$. In view
of the presumed absence of enhanced surface couplings, such a behavior
should be due to an ``effective'' {\em ordering surface field\/}
$g_1=g_1(K,h,h_1)$. Away from $T_c$ and in the disordered phase,
such a field generates a linear response of the local OP
which decays exponentially into the bulk. A glance at (\ref{profiles100})
leads us to anticipate the form
\[
g_1(K,h,h_1) = h_1+m_{\text{dis}}(K,h)\,.
\]
We will derive a formula for $g_1$ identical with the above expression
in Sec.\ \ref{GL}, when we map the lattice model onto a continuum theory.
There it will become clear that $g_1$ is indeed a surface field coupling
to the local OP that enters into a coarse-grained (Ginzburg-Landau)
free-energy functional.


\subsection{Solutions of the nonlinear recursive maps}\label{latticeMFT_num}
The thermodynamically stable solutions of the bulk MF equations correspond
to hyperbolic fixed points of the maps $F^2$ and $G$
(cf.\ Sec.\ \ref{LinearMF}). The stable (unstable) manifold $W_s$ ($W_u$),
or inset (outset), of a hyperbolic fixed point consists of all points
that converge to the fixed point under iteration of the map (inverse map).
In order to solve the MF equations for the semi-infinite system one must
determine the intersections of the stable manifolds with the linear subspaces
defined by the surface boundary conditions (\ref{BC_surf100})
and (\ref{BC_surf110}), i.~e., the line
$\left\{(-h_1,x)^T\big|\,x\in{\Bbb R}\right\}$
and the hyperplane $\left\{(-2h_1,-2h_1,x,y)^T\big|\,x,y\in{\Bbb R}\right\}$.
The magnetization densities of the first layer can be read off
from the intersection points $\bbox{v}_1=(-h_1,m_1)^T$ and
$\bbox{w}_1=(-2h_1,-2h_1,m_{1,\alpha},m_{1,\beta})^T$.
The complete magnetization profile follows from the trajectories passing
through $\bbox{v}_1$ and $\bbox{w}_1$.\cite{numerics} Below $T_c$, infinitely
many intersections exist and one has to resort to the original variational
principle to find the equilibrium profile minimizing the free-energy functional
(\ref{MF_fctl}).

\subsubsection{(100) surface, $T\ge T_c$}
\begin{figure}[t]
\epsfxsize=8.6cm
\centerline{\epsffile{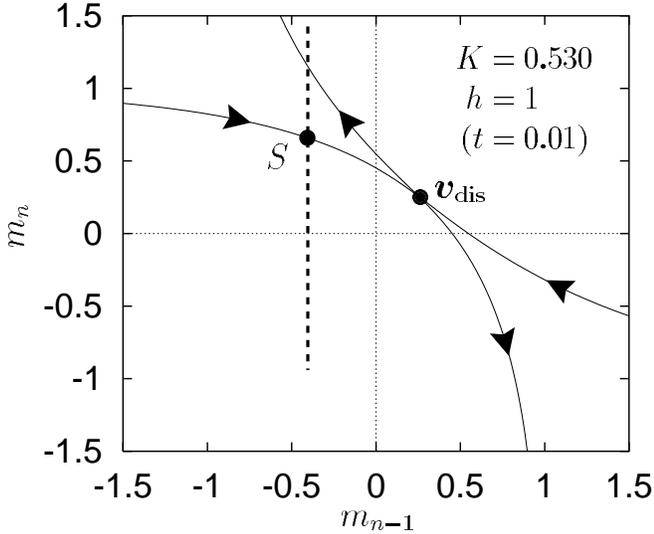}}
\vspace{2mm}
\caption
{
Invariant manifolds (inset and outset) of the hyperbolic fixed point
$\protect\bbox{v}_{\text{dis}}$. The direction of the flow under iteration
of the map $F$ is indicated by arrows. The dashed line represents the boundary
condition for $h_1=0.4$.
}
\label{Fig_invMfld100a}
\end{figure}
\begin{figure}[t]
\epsfxsize=8cm
\epsffile{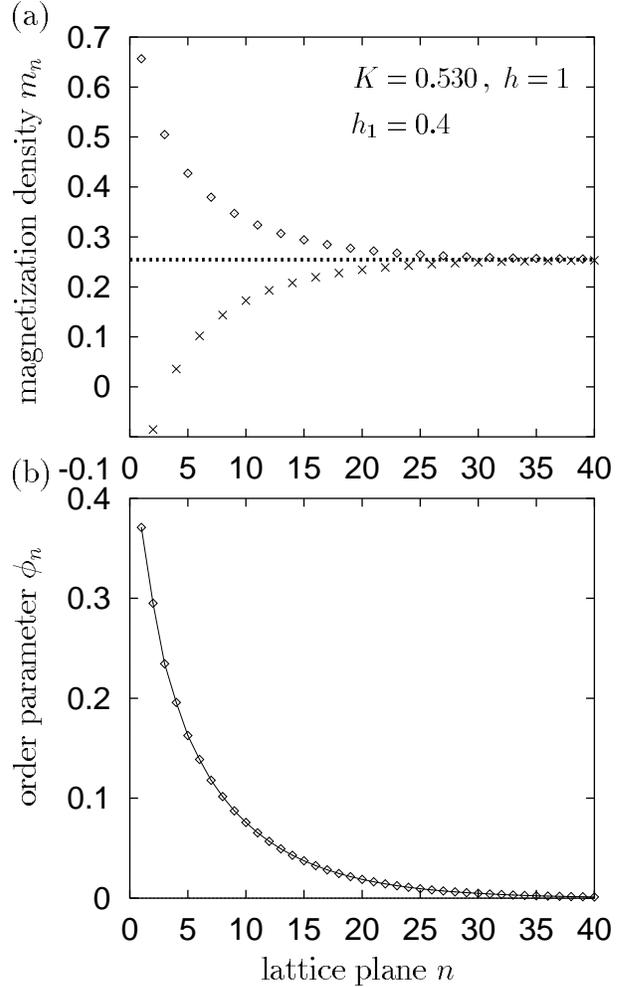}
\vspace{2mm}
\caption
{
(a) Magnetization profile $m_n$ for the (100) surface above $T_c$,
as obtained from the intersection $S$ of Fig.\ \protect\ref{Fig_invMfld100a}.
Odd and even planes $n$ belong to sublattice $\alpha$ ($\Diamond$)
and $\beta$ ($\times$), respectively. Note that $m_n$
oscillates about the bulk value $m_{\text{dis}}$ (dashed line).
(b) OP profile, defined by (\protect\ref{OPasym}).
For convenience, the discrete points have been joined by lines.
}
\label{Fig_prof100_T>Tc}
\end{figure}

In the disordered phase, the only fixed point of $F^2$ (as well as of $F$) is
$\bbox{v}_{\text{dis}}=(m_{\text{dis}},m_{\text{dis}})^T$
[Eq.\ (\ref{FP_dis})]. Fig.\ \ref{Fig_invMfld100a} shows a plot
of the invariant manifolds for particular values of $K$ and $h$.
Fig.\ \ref{Fig_prof100_T>Tc} depicts the magnetization and OP profiles
obtained from the intersection with the boundary condition for $h_1=0.4$. Since
$R\circ F=F^{-1}\circ R$ [cf.\ Eq.\ (\ref{refl_symm})], the inset and outset
are mapped onto each other under reflection at the line $x=y$.

The picture of the invariant manifolds does not change qualitatively
as one varies the parameters $K$ and $h$. From the eigenvector $\bbox{l}_2$
[Eq.\ (\ref{evecs100})] one infers that the slope of the inset
at $\bbox{v}_{\text{dis}}$ vanishes in the high-temperature limit ($K\to0$),
whereas it approaches $-1$ for $K\uparrow K_c$ ($T\downarrow T_c$).

The upshot is that we find a unique intersection of the boundary condition
with the stable manifold. In particular, we obtain a nonvanishing order
parameter profile at any temperature $T>T_c$. This conforms with the idea
that the OP profile is due to an ordering surface field.

An exceptional case occurs if the boundary condition exactly hits
the fixed point $\bbox{v}_{\text{dis}}$, so that $m_n\equiv m_{\text{dis}}$
and $\phi_n\equiv0$. However, except in the case $h=h_1=0$,
this can only be achieved for a special temperature $T=T_0$
(at fixed $h$ and $h_1$). For $T\neq T_0$ the OP profile is still nonzero.

\subsubsection{(110) surface, $T\ge T_c$}\label{110_T>Tc}
\begin{figure}[tb]
\epsfxsize=8.6cm
\epsffile{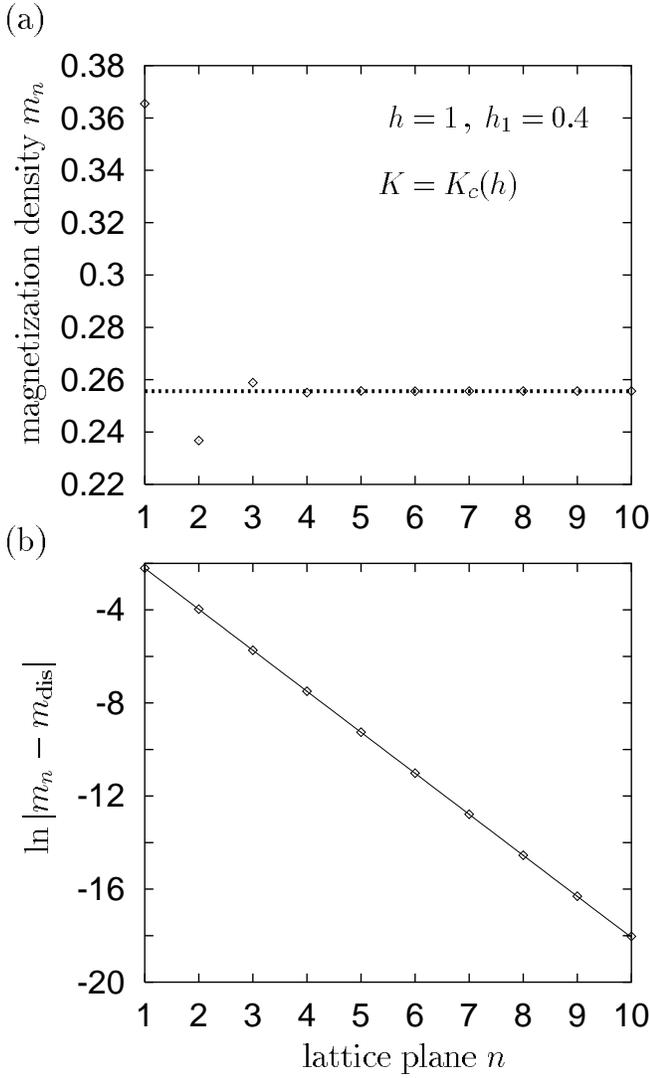}
\vspace{2mm}
\caption
{
(a) Magnetization profile at $T=T_c$ for the (110) surface. The dashed line
represents the bulk value $m_{\text{dis}}$.
(b) Exponential decay of the profile. The solid curve is a straight line
with slope $\tilde{\xi}^{-1}$ [Eq.\ (\protect\ref{length_scale110})].
}
\label{Fig_prof110}
\end{figure}

As we have seen in Sec.\ \ref{Intro}, this type of surface
is symmetry-preserving and the Hamiltonian (\ref{Hamiltonian})
is exactly symmetric with respect to interchanging $\alpha$ and $\beta$-sites.
Since we precluded the possibility of supercritically enhanced surface bonds,
a spontaneous breakdown of this symmetry is ruled out for $T\ge T_c$.
Therefore the solutions that minimize the free energy in this temperature
regime fulfill $m_{n,\alpha}=m_{n,\beta}\equiv m_n$, and $G$ acts
in a 2D subset $\left\{(x,x,y,y)^T\big|\,x\in{\Bbb R}\,,\;|y|<1\right\}$
of ${\Bbb R}^4$. The picture of the invariant manifolds looks similar
to Fig.\ \ref{Fig_invMfld100a}. The magnetization profile at the bulk critical
point $K=K_c(h)$ (for $h=1$) is depicted in Fig.\ \ref{Fig_prof110}.
As in the case of the (100) orientation, the layer magnetization densities
oscillate about the bulk value $m_{\text{dis}}$ due to the antiferromagnetic
coupling between neighboring planes. However, the decay length remains
on the order of the lattice constant even at $T=T_c$,
(cf.\ Sec.\ \ref{LinearMF}). The numerical profile decays exponentially
on a length scale that agrees well with the value (\ref{length_scale110}).

\subsubsection{(100) surface, $T<T_c$}
\begin{figure}[b]
\epsfxsize=8.6cm
\epsffile{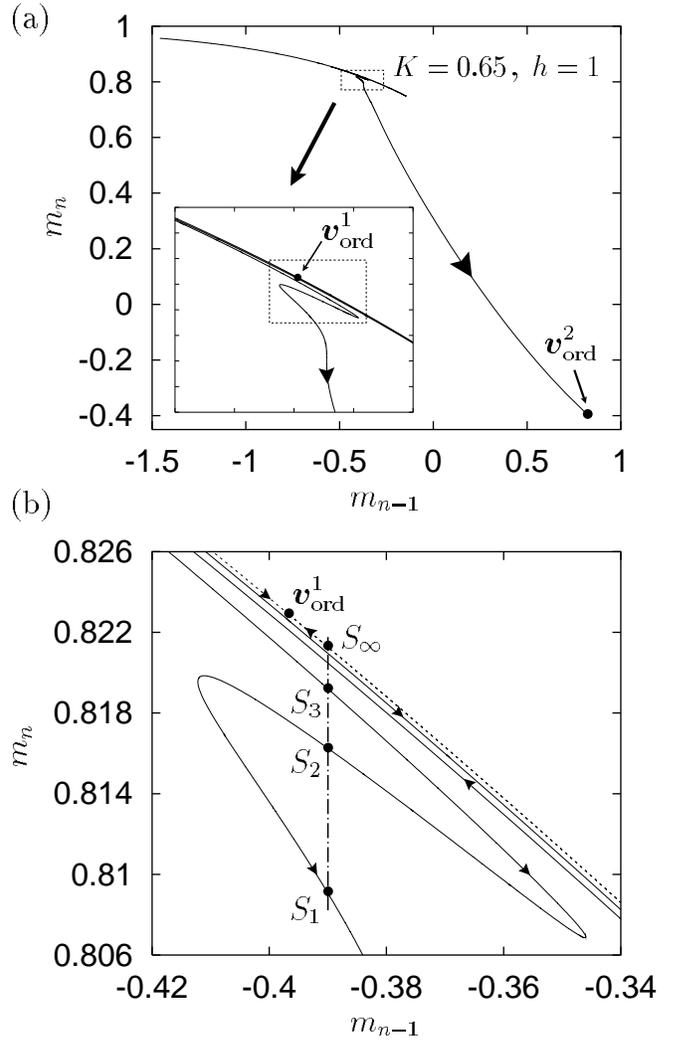}
\vspace{2mm}
\caption
{
(a) Part of the stable manifold of the fixed point
$\protect\bbox{v}_{\text{ord}}^2$. The inset shows the first few loops
in the vicinity of $\protect\bbox{v}_{\text{ord}}^1$. (b) Magnification
of the region marked in the inset of (a). The dashed-dotted
line is the boundary condition for $h_1=0.39$, which intersects
the stable manifold at an infinite number of points $S_1$, $S_2$, $\ldots$.
The latter accumulate at the point $S_\infty$ on the stable manifold
of $\protect\bbox{v}_{\text{ord}}^1$ (dashed line).
}
\label{Fig_invMfld100b}
\end{figure}
On crossing the critical line in the MF phase diagram
(Fig.\ \ref{Fig_phdiagr}), the map $F$ undergoes a period-doubling bifurcation
(Sec.\ \ref{LinearMF}), and the picture of the invariant manifolds
changes qualitatively. The fixed point $\bbox{v}_{\text{dis}}$ becomes elliptic
and looses any inset and outset. The inset (outset)
of any one of the hyperbolic fixed points $\bbox{v}_{\text{ord}}^{1,2}$
cannot intersect itself but will generically intersect the outset (inset)
of the same fixed point, as well as the outsets (insets)
of all other fixed points, at an infinite number of so-called homoclinic
and heteroclinic points.\cite{Arrowsmith} The invariant manifolds oscillate
wildly in the vicinity of the fixed points (cf.\ Fig.\ \ref{Fig_invMfld100b}),
giving rise to the phenomenon of ``chaotic entanglement''.\cite{Arrowsmith}
As a consequence, {\em infinitely\/} many solutions of the lattice MF
equations for the semi-infinite system exist, and one has to resort
to the original variational principle (cf.\ Sec.\ \ref{MFTasNLM})
in order to decide which solution corresponds to the true equilibrium profile.
The stable and unstable manifolds $W_s$ and $W_u$ of $\bbox{v}_{\text{ord}}^1$
and $\bbox{v}_{\text{ord}}^2$ are determined uniquely
given only one of them, say $W_s\!\left(\bbox{v}_{\text{ord}}^1\right)$.
In fact, from the symmetry property (\ref{refl_symm}) one concludes that
\BAA
W_u\left(\bbox{v}_{\text{ord}}^1\right) &=&
F\circ R\biglb(W_s\!\left(\bbox{v}_{\text{ord}}^1\right)\bigrb)\,,\\
W_s\left(\bbox{v}_{\text{ord}}^2\right) &=&
F\biglb(W_s\!\left(\bbox{v}_{\text{ord}}^1\right)\bigrb)\,,\\
W_u\left(\bbox{v}_{\text{ord}}^2\right) &=&
R\biglb(W_s\!\left(\bbox{v}_{\text{ord}}^1\right)\bigrb)\,.
\EAA

\begin{figure}[t]
\epsfxsize=8.6cm
\epsffile{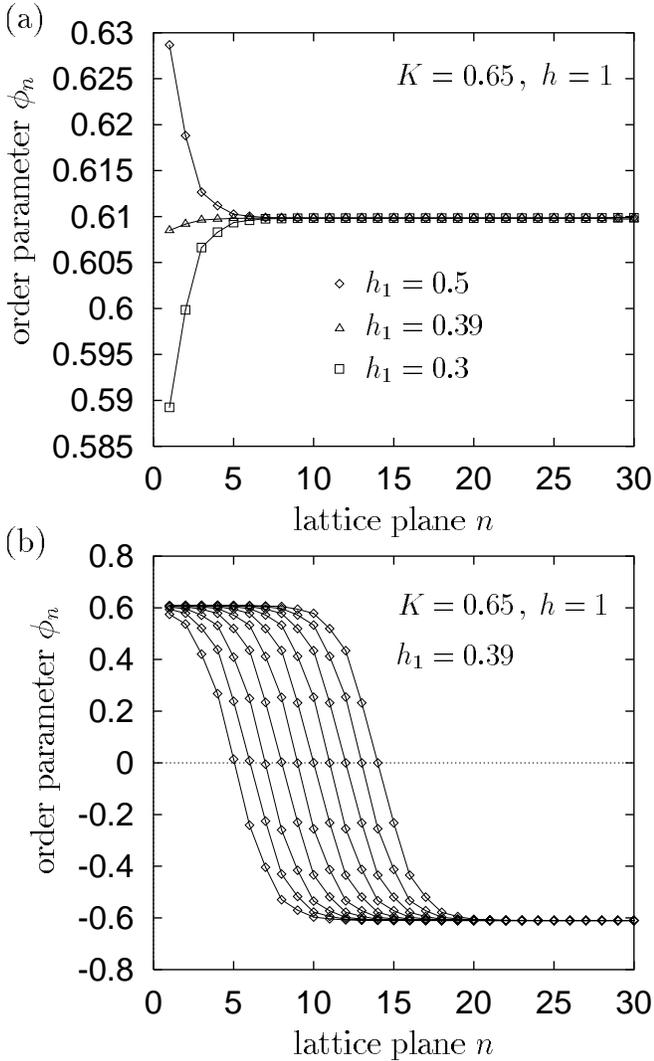}
\vspace{2mm}
\caption
{
(a) OP profiles below $T_c$ for the (100) surface.
The corresponding trajectories converge to the fixed point
$\protect\bbox{v}_{\text{ord}}^1$
(cf.\ Fig.\ \protect\ref{Fig_invMfld100b}b).
The middle profile ($h_1=0.39$) belongs to the intersection $S_\infty$.
(b) OP profiles describing an antiphase boundary. The trajectories
converge to $\protect\bbox{v}_{\text{ord}}^2$.
The leftmost profile belongs to the intersection $S_1$, the next one to $S_2$,
etc.\ (Fig.\ \protect\ref{Fig_invMfld100b}b). An infinite number of such
solutions exists, the first ten of which are shown here.
}
\label{Fig_prof100b}
\end{figure}
Typical minimum-free-energy profiles are shown in Fig.\ \ref{Fig_prof100b}a.
The corresponding trajectories converge to the fixed point $\bbox{v}_1$
describing the bulk phase $\phi_b>0$.
If one imposes the bulk boundary condition $\phi_n\to-\phi_b<0$ ($n\to\infty$),
so that the solutions converge to $\bbox{v}_2$,
one obtains OP profiles exhibiting an {\em antiphase boundary},
i.~e., an interface between the two phases $\pm\phi_b$
(Fig.\ \ref{Fig_prof100b}b).
These solutions always yield a higher free energy than the profiles
of Fig.\ \ref{Fig_prof100b}a. For the chosen parameter values,
the free energy of the profiles is found to {\em increase\/}
as the position of the interface moves into the bulk, so that the leftmost
profile represents the equilibrium solution for this type
of boundary conditions. By analogy with wetting phenomena,\cite{Dietrich-DL12}
we may say that the surface is ``nonwet'', i.~e., the interface has a finite
(microscopic) distance from the surface. If $h_1$ is increased,
so that the ``effective'' ordering surface field $g_1$ favoring the bulk phase
$\phi_b>0$ becomes strong enough, the free energy of the profiles eventually
{\em decreases\/} and the surface is ``wet''. (E.~g., in the situation
depicted in Fig.\ \ref{Fig_prof100b}b this happens if one chooses
$h_1=0.4$.) In Ref.\ \onlinecite{LDrD} the wetting phase
diagram in the space of thermodynamic parameters $K$, $h$, and $h_1$
is calculated using the continuum model derived in the next section.


\section{Ginzburg-Landau theory}\label{GL}
In this section, the aim is to derive and critically examine
a Ginzburg-Landau model for the semi-infinite alloy with a (100) surface.
In particular, we want to show that the loss
of the $\alpha\leftrightarrow\beta$ sublattice symmetry
(cf.\ Sec.\ \ref{Intro}) leads, in a continuum description,
to a {\em symmetry-breaking boundary condition\/} for the OP profile $\phi(z)$,
\[
\dot{\phi}(0) = \frac{1}{\lambda}\phi(0) - \frac{g_1}{C}\,,
\]
where $z=0$ corresponds to the surface plane ($n=1$)
and the dot denotes differentiation with respect to $z$.
Such a boundary condition is familiar from the phenomenological theory
of surface critical phenomena.\cite{Binder-DL8} The parameter $C>0$
is the coefficient of the gradient term of the Ginzburg-Landau functional.
The {\em extrapolation length\/} $\lambda$ should be positive
owing to the absence of enhanced surface bonds. Thus the persistence
of surface order for $T\ge T_c$ originates solely from the ``effective''
{\em ordering surface field\/} $g_1\neq0$. We will determine the dependence
of $g_1=g_1(K,h,h_1)$ on the reduced coupling constant $K$ and
the fields $h$ and $h_1$.

No such ordering surface field emerges in the case of the symmetry-preserving
(110) surface. However, the Ginzburg-Landau functional depends
on an additional spatially varying nonordering density
to which the surface field $h_1$ couples. We defer the derivation
of the corresponding continuum model to a subsequent paper,\cite{L} where
nonordering densities and the construction of suitable multicomponent
Ginzburg-Landau theories will be treated from a more general perspective.

Upon approaching the bulk critical point in the presence of an ordering
surface field $g_1\neq0$, the semi-infinite system undergoes the {\em normal\/}
transition, which exhibits critical singularities that are distinct
from those of the {\em ordinary\/} transition. The latter occurs for $g_1=0$
and subcritical surface enhancement ($\lambda>0$).\cite{g1small}
In order to confirm that the continuum theory derived in Sec.\ \ref{cont_limit}
correctly describes the asymptotic critical behavior of the lattice model,
we shall draw a detailed numerical comparison with the solutions
of the lattice MF equations and clearly identify the singular behavior
of the normal transition for generic values of $h$ and $h_1$
(Sec.\ \ref{normal}). By tuning $h$ and $h_1$, one can achieve that $g_1=0$
at $T=T_c$. Then the singularities of the ordinary transition are recovered,
although the OP at the surface is nonzero for $T>T_c$ because of $g_1\neq0$
for $T\neq T_c$ (Sec.\ \ref{ordinary}).

\subsection{Derivation of continuum model}\label{cont_limit}
One has to be careful to define the local OP $\phi_n$
in such a way that the {\em space-inversion symmetry\/} of the lattice MF
equations (\ref{MF100}) survives the continuum limit. If the magnetization
profile $m_n$, $n=1,2,\ldots$, is a solution of (\ref{MF100}), so is the
profile $\tilde{m}_{-n}\equiv m_{n+2}$ obtained by reflection at the surface
plane $n=1$. If the definition of $\phi_n$ respects this symmetry,
the differential equation for the continuum profile $\phi(z)$
should be invariant under $\phi(z)\to\phi(-z)$. However, the definition
(\ref{OPasym}) distinguishes one direction along the [100] axis
and violates space-inversion symmetry explicitly. As a consequence,
first-order derivatives of $\phi(z)$ appear in the Ginzburg-Landau equations,
corresponding to {\em linear\/} derivative terms in the free-energy
functional.\cite{Schmid} Such terms render
the functional unbounded from below and thus preclude it from serving
as a Landau-Ginzburg-Wilson Hamiltonian of a field theory.

To avoid these difficulties, we adopt an alternate definition of the local OP
treating the {\em two\/} neighboring layers of lattice plane $n$
on an equal footing:
\BE\label{OPsym}
\phi_n\equiv\frac{1}{2}(-1)^n\left[\frac{1}{2}(m_{n-1}+m_{n+1})-m_n\right]\,.
\EE
This definition complies with space-inversion symmetry
and coincides with (\ref{OPasym}) in the bulk of the system.

Substituting $m_{n+1}+m_{n-1}$ from (\ref{MF100}) into (\ref{OPsym}) we obtain
\BE\label{OPloc}
\phi_n=(-1)^{n+1} \varphi(m_n)\,,
\EE
where the function
\[
\varphi(x)\equiv-\frac{h}{4}+\frac{x}{2}+\frac{1}{4K}\tanh^{-1}x
\]
is strictly monotonous and thus invertible. We denote the inverse of $\varphi$
by ${\cal M}$, so that from (\ref{OPloc})
\BE
m_n =
{\cal M}\biglb((-1)^{n+1}\phi_n\bigrb)\,.\eqnum{\ref{OPloc}$^\prime$}
\EE
The continuum limit of the lattice MF equations (\ref{MF100})
leads to the Ginzburg-Landau equation (see Appendix \ref{App_GL})
\BE\label{GLbulk}
C\ddot{\phi} = 4\phi+2{\cal M}(-\phi)-2{\cal M}(\phi)\,,
\EE
where $C\equiv{\cal M}^\prime(0)$ [Eq.\ (\ref{M_1})]. The MF equation
at the surface (\ref{MF100_surf}) entails the boundary condition
(see Appendix \ref{App_BC})
\BE\label{GLsurf}
C\dot{\phi}(0) = -h_1-{\cal M}\bbox{(}\!-\!\phi(0)\bbox{)}\,.
\EE
For a spatially homogeneous system, (\ref{GLbulk}) is identical to the equation
for the OP following from the bulk MF equations (\ref{MFbulk}$^\prime$).
In fact, (\ref{MFbulk}$^\prime$) can be written as
\[
\phi = \varphi(m+\phi)\,,\quad -\phi = \varphi(m-\phi)\,.
\]
Operating with $\cal M$ on both sides of the above equations and eliminating
$m$, one arrives at (\ref{GLbulk}) with $\ddot{\phi}\equiv0$.

We denote the bulk Landau free-energy density by $f_b(m_\alpha,m_\beta)$
[Eq.\ (\ref{fb})]. Since
\BAA
\partial_\alpha f_b(m_\alpha,m_\beta) &=&
2\varphi(m_\alpha)-(m_\alpha-m_\beta)\,,\\
\partial_\beta f_b(m_\alpha,m_\beta) &=&
2\varphi(m_\beta)+(m_\alpha-m_\beta)\,,
\EAA
where $\partial_\mu\equiv\frac{\partial}{\partial m_\mu}$, $\mu=\alpha,\beta$,
Eqs.\ (\ref{GLbulk}) and (\ref{GLsurf}) can be rewritten as
\BA
C\ddot{\phi} &=& V^\prime(\phi)\,,\eqnum{\ref{GLbulk}$^\prime$}\\
C\dot{\phi}(0) &=& f_s^\prime(\phi)\,,\eqnum{\ref{GLsurf}$^\prime$}
\EA
where
\BA
V(\phi) &=& \frac{1}{2}[{\cal M}(\phi)-{\cal M}(-\phi)-2\phi]^2\nonumber\\
\label{V}
& & {}+ f_b\bbox{(}{\cal M}(\phi),{\cal M}(-\phi)\bbox{)}
- f_b\bbox{(}{\cal M}(0),{\cal M}(0)\bbox{)}\,,\\
f_s(\phi) &=& -[h_1+{\cal M}(-\phi)]\phi
+\frac{1}{4}f_b\bbox{(}{\cal M}(-\phi),{\cal M}(-\phi)\bbox{)}\,.\nonumber\\
& & {}-\frac{1}{4}f_b\bbox{(}{\cal M}(0),{\cal M}(0)\bbox{)}\,.\label{fs}
\EA
Thus the Ginzburg-Landau equations (\ref{GLbulk}) and (\ref{GLsurf})
follow from the variation of the free-energy functional
\BE\label{GLfctl}
{\cal F}[\phi] =
\int_0^\infty dz\,\left[\frac{C}{2}\dot{\phi}^2+V(\phi)\right]
+ f_s\bbox{(}\phi(0)\bbox{)}\,.
\EE
Expansion of $V(\phi)$ yields the usual $\phi^4$-form
\BE\label{V_exp}
V(\phi) = \frac{A}{2}\phi^2 + \frac{B}{4}\phi^4
+ {\cal O}\left(\phi^6\right)\,,
\EE
where $A=4(1-M_1)$ and $B=-4M_3$ [Eqs.\ (\ref{M_i})].
With the aid of (\ref{mdis_exp}), the leading temperature dependence
of the Landau coefficients is found to be
\BA
A &=& A_1t + {\cal O}\left(t^2\right)\,,\eqnum{\ref{V_exp}a}\\
B &=& B_0 + {\cal O}(t)\,,\eqnum{\ref{V_exp}b}\\
C &=& 1 + {\cal O}(t)\,.\eqnum{\ref{V_exp}c}
\EA
where
\BAA
A_1 &=& A_1(h) = 2\{1-K_c(h)m_c(h)[h-2m_c(h)]\}\,,\\
B_0 &=& B_0(h) = \frac{8}{3}K_c(h)^2\,.
\EAA
Likewise, the Landau expansion of $f_s(\phi)$ reads
\BE\label{fs_exp}
f_s(\phi) = - g_1\phi + \frac{C}{2\lambda}\phi^2
+ {\cal O}\left(\phi^3\right)\,,
\EE
where
\BA
g_1(K,h,h_1) &=& h_1 + m_{\text{dis}}(K,h)\,,\eqnum{\ref{fs_exp}a}\\
\lambda &=& 1\,.\eqnum{\ref{fs_exp}b}
\EA
Thus we obtain a {\em positive\/} extrapolation length $\lambda$
and an ordering surface field $g_1\neq0$, as anticipated above.

\subsection{Comparison with lattice MF theory:
generic (nonideal) stoichiometry}\label{normal}
If $h$ and $h_1$ take generic values (nonideal bulk and surface stoichiometry),
$g_1=g_1(K,h,h_1)$ is nonzero at $T=T_c$
and gives rise to an OP profile decaying according
to a power law. For $t=0$, (\ref{GLbulk}$^\prime$) becomes
\BE\label{GL_Tc}
\ddot{\phi} = B_0\phi^3\,.
\EE
The neglected higher powers of $\phi$ do not affect the asymptotic behavior
of $\phi(z)$ as $z\to\infty$. The solution of (\ref{GL_Tc})
satisfying $\phi(z)\to0$ as $z\to\infty$ reads
\BE
\phi(z)=\frac{P_0}{z+z_0}\,,\eqnum{\ref{GL_Tc}a}
\EE
with the amplitude
\BE
P_0(h) = \pm\sqrt{\frac{2}{B_0(h)}}=\pm\frac{\sqrt{3}}{2K_c(h)}\,.
\eqnum{\ref{GL_Tc}b}
\EE
The signs refer to positive and negative $g_1$, respectively.
The integration constant $z_0$ follows by inserting (\ref{GL_Tc}a)
into the boundary condition (\ref{GLsurf}).
An algebraic decay $\phi(z)\sim z^{-x_\phi}$ of the OP profile at $T=T_c$,
where $x_\phi$ is the scaling dimension of the OP,
is characteristic of the normal transition.
Generally, $x_\phi$ can be expressed by bulk critical exponents as
$x_\phi=\beta/\nu$. Within MF theory, $\beta=\nu=1/2$, so that $x_\phi=1$.
\begin{figure}[t]
\epsfxsize=8.6cm
\epsffile{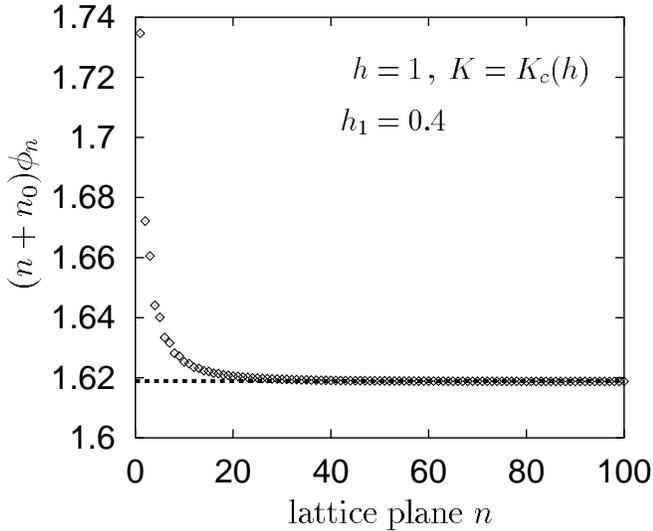}
\vspace{2mm}
\caption{Scaling of the critical OP profile $\phi_n$ [defined by
(\protect\ref{OPsym})]. The expected power law decay has been factored out.
The parameter $n_0$ was fitted optimally. The dashed line is the prediction
(\protect\ref{GL_Tc}b) of the continuum theory.
}
\label{Fig_prof_scaling}
\end{figure}

\begin{figure}
\epsfxsize=8.6cm
\epsffile{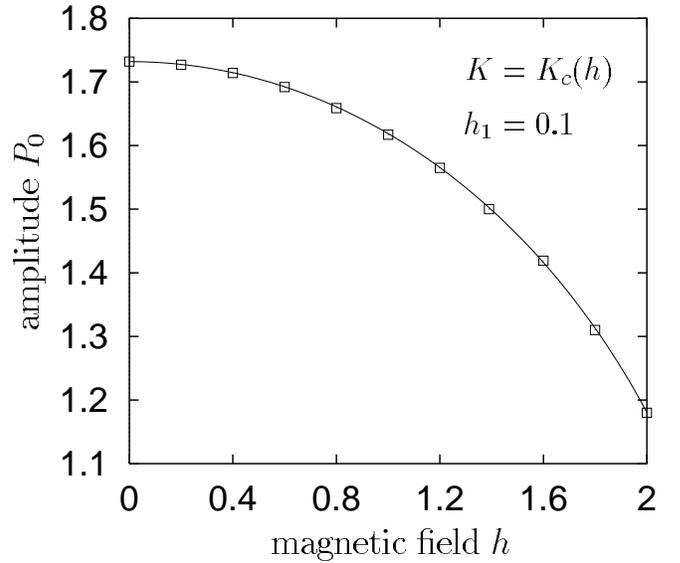}
\vspace{2mm}
\caption
{
Amplitude of the critical OP profile as a function of the bulk field $h$.
The data obtained from the solution of the lattice MF equations ($\Box$)
agree well with the prediction (\protect\ref{GL_Tc}b) of the continuum model
(full line).
}
\label{amplitude}
\end{figure}

In order to check whether the lattice MF profiles exhibit
the above power law decay with the predicted value (\ref{GL_Tc}b)
for the amplitude, we used the ``nonlinear-mapping representation''
of the lattice MF equations (Sec.\ \ref{NLmaps}) to determine the profiles
numerically up to $\simeq1000$ layers from the surface.
Fig.\ \ref{Fig_prof_scaling} demonstrates that
the power law decay is well reproduced. The amplitude extracted
from the numerical fits is in perfect agreement with (\ref{GL_Tc}b)
for a wide range of the bulk field $h$ (Fig.\ \ref{amplitude}).

As a second test of the validity of our continuum model we investigate
the temperature singularity of the surface OP.
We multiply the Ginzburg-Landau equation (\ref{GLbulk}$^\prime$)
by $\dot{\phi}$ and perform the integral from $z=0$ to $z=\infty$
on both sides using the boundary conditions (\ref{GLsurf}$^\prime$)
and $\phi(z)\to0$ as $z\to\infty$. This gives
\BE\label{surfEOS}
-\frac{C}{2}f_s^\prime(\phi_s)^2 + V(\phi_s) = V_\infty\,,
\EE
where $\phi_s\equiv\phi(0)$ and $V_\infty\equiv V(\phi_b)$. One has
\[
V_\infty = \left\{\begin{array}{lr}
0\,, & t>0\,,\\
-\frac{A_1^2}{2B_0}t^2+{\cal O}\left(t^3\right)\,, & t<0\,.
\end{array}\right.
\]
Expanding the coefficients on both sides of the surface equation of state
(\ref{surfEOS}) in powers of $t$,
one recognizes that $\phi_s$ exhibits a discontinuity in the second temperature
derivative due to the non-analyticity of $V_\infty$ at $t=0$,
\BE\label{phis_normal}
\phi_s = \phi_s^{(0)} + \phi_s^{(1)}t + \phi_{s,\pm}^{(2)}t^2 + \ldots\,,
\EE
with $\phi_{s,+}^{(2)}\neq\phi_{s,-}^{(2)}$. This result complies with
the general form of the leading $|t|^{2-\alpha}$ singularity of the surface
OP at the normal and extraordinary transitions.\cite{ampl-normal}
\begin{figure}
\epsfxsize=8.6cm
\epsffile{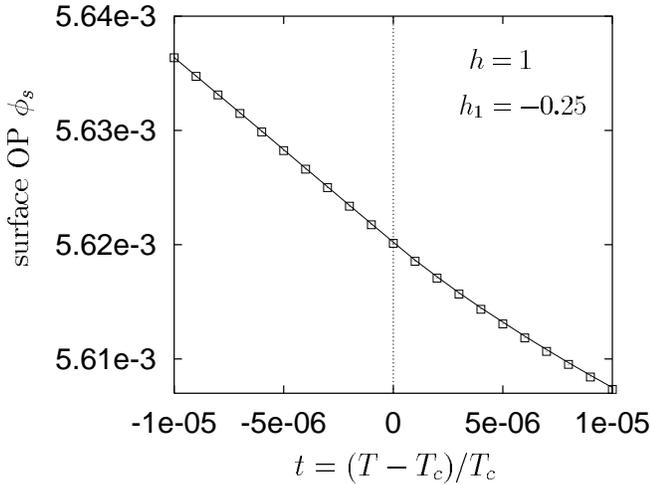}
\vspace{2mm}
\caption{Temperature dependence of the surface OP $\phi_s$
for generic values of $h$, $h_1$, in the immediate vicinity
of the critical temperature. The results of the lattice MF theory ($\Box$)
agree well with the prediction of the continuum model (solid line).
}
\label{Fig_normal}
\end{figure}

Fig.\ \ref{Fig_normal} shows a comparison with the solutions of the lattice MF
equations. While the singularity in the second temperature derivative
is too weak to be detected without considerable numerical effort, the data
are fully consistent with the continuity of the first temperature derivative
at $t=0$ and agree well with the predictions of the continuum theory.
The continuity of the first temperature derivative of $\phi_s$
is incompatible with the ordinary transition and confirms again
that the asymptotic critical behavior falls into the universality class
of the normal transition.

\subsection{Comparison with lattice MF theory: Vanishing ordering surface
field at $T=T_c$}\label{ordinary}
According to (\ref{fs_exp}a), $g_1$ can be made to vanish at $T=T_c$
by choosing $h_1$ such that
\BE\label{g1zero}
h_1=-m_c(h)\,.
\EE
If $h=h_1=0$ (ideal bulk and surface stoichiometry), $g_1\equiv0$ for
{\em all\/} temperatures, and the system clearly displays
ordinary surface critical behavior. If $h\neq0$ and (\ref{g1zero})
is fulfilled, $g_1$ varies linearly with $t$ as $t\to0$. In particular,
{\em the surface OP $\phi_s$ is nonzero for $T>T_c$\/} and vanishes only
in the limit $t\to0\pm$. One may wonder whether such a behavior
is consistent with the ordinary transition where one usually expects
that $\phi_s\equiv0$ for $T\ge T_c$.

To derive the leading temperature singularity of $\phi_s$
consider again Eq.\ (\ref{surfEOS}).
Since $\phi_s=0$ if $t=0$, $\phi_s$ is found to behave asymptotically as
\BE\label{phis_ord}
\phi_s=\phi_{s,^\pm}^{(1)}t + \text{less singular terms}\,,
\EE
with $\phi_{s,+}^{(1)}\neq\phi_{s,-}^{(1)}$.
The ordering surface field (\ref{fs_exp}a) can be expanded as
\BE\label{g1_exp}
g_1(K,h,h_1) = g_1^{(0)}(h,h_1) + g_1^{(1)}(h)t + {\cal O}\left(t^2\right)\,,
\EE
where $g_1^{(0)}(h,h_1)=h_1+m_c(h)=0$ owing to (\ref{g1zero}) and
$g_1^{(1)}(h)=m_{\text{dis}}^{(1)}(h)$ [Eq.\ (\ref{mdis_exp})].
Insertion into (\ref{surfEOS}) yields, to leading order in $t$,
\BE\label{surfEOS-ord}
\left(-g_1^{(1)}+\phi_{s,1}^\pm\right)^2 =
\left\{\begin{array}{l@{\quad,\quad}r}
0 & t>0\,,\\
\frac{A_1^2}{B_0} & t<0\,.
\end{array}
\right.
\EE
Therefore one concludes that \cite{ampl-ord}
\BE\label{ampl_ord}
\phi_{s,1}^+=g_1^{(1)}\,,\quad \phi_{s,1}^-=g_1^{(1)}-\frac{A_1}{\sqrt{B_0}}\,.
\EE
The discontinuity of $\phi_s$ in the {\em first\/} temperature derivative
differs strikingly from the singularity in the second derivative
for generic values of $h$, $h_1$ (Sec.\ \ref{normal}). This result is
in excellent agreement with the numerical solutions of the lattice MF equations
(Fig.\ \ref{Fig_ordinary}). As will be shown below, such a behavior is
precisely what one expects if the leading asymptotic behavior belongs
to the universality class of the {\em ordinary\/} transition. The variation
of the effective ordering surface field $g_1$ with temperature
[cf.\ Eq.\ (\ref{g1_exp})] explains the onset of surface order for $t>0$
(see below).

Let us elucidate the above behavior of the surface OP $\phi_s$ by resort
to a scaling argument. The basic assumption is the existence of a scaling field
$\tilde{g}_1=\tilde{g}_1(K,h,h_1)$ associated with $\phi_s$ and depending
analytically on $K$, $h$, and $h_1$. Of course, $\tilde{g}_1$ will in general
differ from the MF expression (\ref{fs_exp}a). By analogy with (\ref{g1_exp})
we write
\BE\label{g1gen_exp}
\tilde{g}_1(K,h,h_1) = \tilde{g}_1^{(0)}(h,h_1) + \tilde{g}_1^{(1)}(h,h_1)t
+ {\cal O}\left(t^2\right)\,.
\EE
We suppose that $\tilde{g}_1^{(0)}(h,h_1)$ vanishes if $h$ and $h_1$ are chosen
appropriately. This should always be possible since $\phi_s$ is
positive for large positive $h_1$, and becomes negative if one lets
$h_1\to-\infty$, assuming $h$ and $t$ to be fixed. Thus for given $h$
and $t=0$, $\phi_s$ must vanish for a special value of $h_1$, in which case
one clearly has $\tilde{g}_1^{(0)}=0$.
\begin{figure}
\epsfxsize=8.6cm
\epsffile{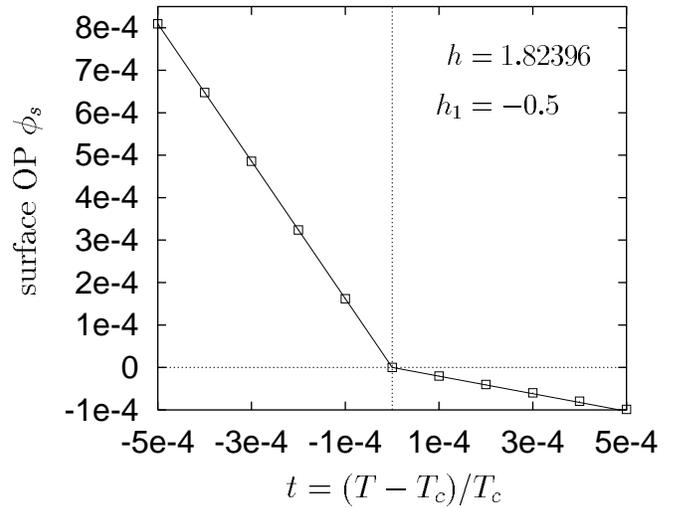}
\vspace{2mm}
\caption
{
Temperature dependence of the surface OP in the case where
the effective ordering surface field $g_1$ vanishes at $T=T_c$.
The discontinuity in the first temperature derivative predicted
by the continuum model (full line) is well confirmed by the numerical solutions
of the lattice MF equations ($\Box$).
}
\label{Fig_ordinary}
\end{figure}

The singular part of the surface free energy $f_s^{\,\text{sing}}$ should take
the standard scaling form \cite{Binder-DL8}
\[
f_s^{\,\text{sing}}(t,\tilde{g}_1) =
M_t^{(d-1)\nu}|t|^{(d-1)\nu}g_{\pm}\left(M_{\tilde{g}_1}M_t^{-\Delta_1}
\tilde{g}_1|t|^{-\Delta_1}\right)\,,
\]
where $\Delta_1\simeq0.48$. All non-universality is embodied
in the metric factors $M_t$ and $M_{\tilde{g}_1}$ associated with the two
relevant scaling fields (at the ordinary transition) $t$ and $\tilde{g}_1$,
while the critical exponents and the scaling functions $g_\pm(\zeta)$
are {\em universal}. The singular part of $\phi_s$ follows by taking
the derivative with respect to $\tilde{g}_1$, i.~e.,
\BE\label{scaling_surfOP}
\phi_s - \phi_s^{\text{(reg)}} = M_t^{\beta_1}M_{\tilde{g}_1}|t|^{\beta_1}
Y_\pm\left(M\tilde{g}_1|t|^{-\Delta_1}\right)\,,
\EE
where $Y_\pm\equiv g_\pm^{\,\prime}$, $M\equiv M_{\tilde{g}_1}M_t^{-\Delta_1}$,
and $\beta_1=(d-1)\nu-\Delta_1$.
The regular contribution $\phi_s^{\text{(reg)}}$ describes, to leading order,
the linear response of $\phi_s$,
\BE
\phi_s^{\text{(reg)}} =
a\tilde{g}_1 + {\cal O}\left(\tilde{g}_1^2\right)\,.
\EE
The importance of such regular terms for the correct identification of surface
critical exponents and scaling functions has been emphasized
in Ref.\ \onlinecite{BM}. The scaling functions $Y_\pm(\zeta)$ are analytic
at $\zeta=0$,
\BA\label{Y}\addtocounter{equation}{1}
Y_+(\zeta) &=& Y_+^{(1)}\zeta + {\cal O}\left(\zeta^2\right)\,,
\eqnum{\theequation a}\\
Y_-(\zeta) &=& Y_-^{(0)} + Y_-^{(1)}\zeta + {\cal O}\left(\zeta^2\right)\,.
\eqnum{\theequation b}
\EA
Using (\ref{g1gen_exp}) with $\tilde{g}_1^{(0)}=0$,
Eqs.\ (\ref{scaling_surfOP})--(\ref{Y}) yield
\[
\phi_s = M_t^{\beta_1}M_{\tilde{g}_1}|t|^{\beta_1}
Y_\pm\left(M\tilde{g}_1^{(1)}t|t|^{-\Delta_1}\right)
+ a\tilde{g}_1^{(1)}t + {\cal O}\left(t^2\right)\,,
\]
where terms of order
$t^p|t|^{-\Delta_1}$, $p=2,3,\ldots$, have been omitted in the argument
of $Y_\pm(\zeta)$. Thus the leading behavior as $t\to0$ becomes
\BE
\phi_s \sim \left\{\begin{array}{l@{\;,\;}l}
a\tilde{g}_1^{(1)}t + M_t^{\beta_1}M_{\tilde{g}_1}M\tilde{g}_1^{(1)}
Y_+^{(1)}t^{1-\gamma_{1,1}}
& t>0\,,\\
M_t^{\beta_1}M_{\tilde{g}_1}Y_-^{(0)}|t|^{\beta_1} & t<0\,,
\end{array}
\right.
\EE
where we used the scaling relation $-\gamma_{1,1}=\beta_1-\Delta_1$.
Since $\gamma_{1,1}<0$ at the ordinary transition,\cite{Binder-DL8}
$\phi_s$ varies linearly with $t$ as $t\to0+$, but vanishes
with the characteristic exponent $\beta_1\simeq0.8$ of the {\em ordinary\/}
transition as $t\to0-$. In MF theory, $\beta_1=1$ and the power singularity
for $t\to0-$ degenerates into an integer power, see Eqs.\ (\ref{phis_ord})
and (\ref{ampl_ord}). That the asymptotic behavior of the ordinary transition
can be obtained by tuning $h$ and $h_1$ has also been demonstrated
by transfer matrix calculations in two dimensions,\cite{Ordinary-2D}
which supplement the results of Ref.\ \onlinecite{Drewitz}.


\section{Summary}\label{Summary}

We have studied the surface critical behavior of bcc binary alloys
undergoing a continuous $B2$-$A2$ order-disorder transition in the bulk.
Clear evidence has been found that symmetry-breaking surfaces,
such as the (100) surface,
generically display the critical behavior of the {\em normal\/} transition,
which belongs to the same universality class as the {\em extraordinary\/}
transition. We have analyzed the lattice MF equations using
the ``nonlinear-mapping'' representation \cite{PW} and achieved a mapping
onto a continuum (Ginzburg-Landau) model. The latter assumes the form
of the standard one-component $\phi^4$-model for semi-infinite systems.
Its crucial feature is the emergence of an ``effective''
{\em ordering surface field\/} $g_1\neq0$, which depends on temperature
and the other parameters of the lattice model and is not present
on a microscopic level. By a detailed comparison with the solutions
of the lattice MF equations the continuum model has been shown to accurately
describe the asymptotic behavior of the lattice model.

In the case of the symmetry-preserving (110) surface
the appearance of an ordering surface field is ruled out by symmetry.
Analysis of the lattice MF equations reveals the existence of an additional
length scale different from the OP correlation length, which describes
the decay of the nonzero magnetization profile above $T_c$. However,
this length stays {\em microscopic\/} even at $T=T_c$ and does not influence
the leading singular behavior, which is characteristic of the {\em ordinary\/}
transition. The construction of a suitable continuum model in this case
is deferred to a subsequent paper,\cite{L} where particular emphasis will be
laid on nonordering densities and the derivation of the corresponding
multicomponent Ginzburg-Landau theories from microscopic models.

\section*{Acknowledgments}
We would like to thank Anja Drewitz for fruitful discussions
and a critical reading of the manuscript. Support by the Deutsche
For\-schungsgemeinschaft (DFG) through Sonderforschungsbereich 237
and the Leibniz program is gratefully acknowledged.


\appendix
\section{Bulk MF equations}\label{App_bulk}
For a spatially homogeneous system with sublattice magnetization densities
$m_\alpha$ and $m_\beta$, the variational free energy (\ref{MF_fctl}) yields
the Landau free-energy density
\BA
\lefteqn{f_b(m_\alpha,m_\beta)\equiv\frac{F_{\text{MFA}}/N}{4|J|}
= m_\alpha m_\beta-\frac{h}{2}(m_\alpha+m_\beta)}\nonumber\\
\label{fb}
& & {}+\frac{1}{2K}\left(
\int_0^{m_\alpha}dx\,\tanh^{-1}x+\int_0^{m_\beta}dx\,\tanh^{-1}x\right)\,,
\EA
where $N$ is the number of lattice sites. The MF equations
\BE\label{MFbulk}
\partial_\alpha f_b = 0\,,\quad\partial_\beta f_b = 0\,,
\EE
where $\partial_\mu\equiv\frac{\partial}{\partial m_\mu}$,
$\mu=\alpha,\beta$, read
\BA
\tanh^{-1}m_\alpha &=& K(h-2m_\beta)\,,\eqnum{\ref{MFbulk}a}\\
\tanh^{-1}m_\beta &=& K(h-2m_\alpha)\,.\eqnum{\ref{MFbulk}b}
\EA
For the following it is useful to define
\BE\label{uab}
u_\alpha\equiv u(m_\alpha)\,,\quad u_\beta\equiv u(m_\beta)\,,
\EE
where
\BE
u(x)\equiv\frac{1}{2K}\frac{1}{1-x^2}\,.
\EE
A solution $(m_\alpha,m_\beta)$ of (\ref{MFbulk}) minimizes $f_b$
and is thus thermodynamically stable if the matrix
\[
\left(\begin{array}{cc}
\partial^2_\alpha f_b & \partial_\alpha\partial_\beta f_b\\
\partial_\alpha\partial_\beta f_b & \partial^2_\beta f_b
\end{array}\right) =
\left(\begin{array}{cc}
u_\alpha & 1\\ 1 & u_\beta
\end{array}\right)
\]
is positive definite, i.~e.,
\BE\label{TDstable}
u_\alpha u_\beta>1\,.
\EE
If one writes $m_\alpha=m+\phi$ and $m_{\beta}=m-\phi$ as in (\ref{mphi}),
the MF equations (\ref{MFbulk}) take the form
\BA
\tanh^{-1}(m+\phi) &=& K(h-2m) + 2K\phi\,,\eqnum{\ref{MFbulk}a$^\prime$}\\
\tanh^{-1}(m-\phi) &=& K(h-2m) - 2K\phi\,.\eqnum{\ref{MFbulk}b$^\prime$}
\EA
The disordered state ($m=m_{\text{dis}}$, $\phi=0$) satisfies
\BE\label{mdis}
\tanh^{-1}m_{\text{dis}} = K(h-2m_{\text{dis}})\,.
\EE
This equation has a unique solution $m_{\text{dis}}=m_{\text{dis}}(K,h)$,
which may be expanded in powers of the reduced temperature $t=(K_c-K)/K$ as
\BE\label{mdis_exp}
m_{\text{dis}} = m_c + m_{\text{dis}}^{(1)}t + {\cal O}\left(t^2\right)\,,
\EE
where $m_c=m_c(h)$ is the magnetization at $T=T_c$, and
\BE
m_{\text{dis}}^{(1)} = m_{\text{dis}}^{(1)}(h)
= -\frac{h-2m_c(h)}{4}\,.\eqnum{\ref{mdis_exp}a}
\EE
The disordered state is thermodynamically stable only if $u(m_{\text{dis}})>1$
[cf.\ Eq.\ (\ref{TDstable})]. The phase transition occurs when
$u(m_{\text{dis}})=1$ [cf.\ Eq.\ (\ref{Kc})]. Two new minima of $f_b$
describing the ordered phases ($m=m_{\text{ord}},\phi=\pm\phi_b$) emerge
if $u(m_{\text{dis}})<1$. The asymptotic behavior following from
(\ref{MFbulk}$^\prime$) is found to be
\BE\label{mphi_exp}
\begin{array}{r@{\;=\;}l}
\D m_{\text{ord}} & \D m_c + m_{\text{ord}}^{(1)}t
+ {\cal O}\left(t^2\right)\,,\\
\D \phi_b & \D \phi_0 |t|^{1/2} + {\cal O}\left(|t|^{3/2}\right)\,,
\end{array}
\EE
where
\BA
m_{\text{ord}}^{(1)} &=& m_{\text{ord}}^{(1)}(h)
= m_{\text{dis}}^{(1)}(h) + K_c(h)m_c(h)\phi_0(h)^2\,,\eqnum{\ref{mphi_exp}a}\\
\phi_0 &=& \phi_0(h) = \frac{\sqrt{3}}{2K_c(h)}
\sqrt{1-K_c(h)m_c(h)[h-2m_c(h)]}\,.\nonumber\\
\eqnum{\ref{mphi_exp}b}
\EA

\section{Ginzburg-Landau equation}\label{App_GL}
Using (\ref{OPloc}) and (\ref{OPloc}$^\prime$), we may rewrite the MF equations
(\ref{MF100}) as
\BE\label{MF_phin}
\begin{array}{l}
\D {\cal M}\biglb((-1)^n\phi_{n-1}\bigrb)
+{\cal M}\biglb((-1)^n\phi_{n+1}\bigrb)\\
\D = 2{\cal M}\biglb((-1)^{n+1}\phi_n\bigrb)+4(-1)^n\phi_n\,.
\end{array}
\EE
In the continuum limit, one replaces $\phi_n$ by a smooth profile $\phi(z)$
defined for all $z\ge0$, with the original layers located at $z_n=n-1$.
Assuming that the OP varies slowly on the scale
of the layer spacing, we approximate
\BE\label{cont_approx}
\begin{array}{l}
\D{\cal M}\biglb((-1)^n\phi_{n-1}\bigrb)
+{\cal M}\biglb((-1)^n\phi_{n+1}\bigrb)\\
\D\simeq 2{\cal M}\biglb((-1)^n\phi(z_n)\bigrb)
+(-1)^n{\cal M}^\prime(0)\ddot{\phi}(z_n)\,,
\end{array}
\EE
where the dot denotes differentiation with respect to $z$ and terms of order
$\phi\ddot{\phi}$ and $\dot{\phi}^2$
have been discarded. Substitution of (\ref{cont_approx}) into (\ref{MF_phin})
leads to the Ginzburg-Landau equation (\ref{GLbulk}).

Since $\varphi(m_{\text{dis}})=0$, i.~e., ${\cal M}(0)=m_{\text{dis}}$,
${\cal M}(\phi)$ may be expanded as
\BE\label{M_exp}
{\cal M}(\phi) = m_{\text{dis}} + M_1\phi + M_2\phi^2 + M_3\phi^3
+ {\cal O}\left(\phi^4\right)\,,
\EE
where
\BA\label{M_i}\addtocounter{equation}{1}
M_1 &=& {\cal M}^\prime(0) =
2[1+u(m_{\text{dis}})]^{-1}\,,\eqnum{\theequation a}\label{M_1}\\
M_2 &=& -Km_{\text{dis}}u(m_{\text{dis}})^2M_1^3\,,\eqnum{\theequation b}\\
M_3 &=& \frac{1}{6}\Bigl\{
3M_1\left[2Km_{\text{dis}}u(m_{\text{dis}})^2\right]^2\nonumber\\
& & {}-(2K)^2\left(1+3m_{\text{dis}}^2\right)u(m_{\text{dis}})^3\Bigr\}
M_1^4\,.\eqnum{\theequation c}
\EA

\section{Boundary condition}\label{App_BC}
By analogy with (\ref{MF_phin}), the MF equation at the surface
(\ref{MF100_surf}) can be written as
\BE\label{MF_phi1}
-h_1 + {\cal M}(-\phi_2) = 2{\cal M}(\phi_1)-4\phi_1\,.
\EE
The continuum approximation (\ref{cont_approx}) now reads
\BE\label{cont_approx_surf}
{\cal M}(-\phi_2)\simeq{\cal M}\bbox{(}-\phi(0)\bbox{)}
-{\cal M}^\prime(0)\left[\dot{\phi}(0)+\frac{1}{2}\ddot{\phi}(0)\right]\,.
\EE
Inserting (\ref{cont_approx_surf}) into (\ref{MF_phi1}) and requiring that
the Ginzburg-Landau equation (\ref{GLbulk}) be also valid at $z=0$, we obtain
\[
h_1+{\cal M}\bbox{(}-\phi(0)\bbox{)}+{\cal M}^\prime(0)\dot{\phi}(0) =
\frac{1}{2}{\cal M}^\prime(0)\ddot{\phi}(0)\,.
\]
The above equation reduces to the boundary condition (\ref{GLsurf}),
if second-order derivatives are neglected.



\begin{references}
\bibitem{MC}
{
The width of the asymptotic region can be estimated by requiring that
the scaling variable $g_1|t|^{-\Delta_1}$ be on the order of one,
where $t=(T-T_c)/T_c$ and $\Delta_1\simeq 0.48$. If we estimate the magnitude
of $g_1$ using the MF expression derived below [Eq.\ (\ref{fs_exp}a)],
we conclude that normal critical behavior should be observed
if $|t|\lesssim0.06$. However, the closest approach
to $T_c$ in the simulations of Ref.\ \onlinecite{Schmid}
was $|t|\simeq 0.14$.
}
\bibitem{bulkphdiagr}
{
The reentrant behavior for $|h|>2$, i.~e.\ the possibility of passing
from the ordered to the disordered phase upon lowering the temperature
at fixed $h$, is not confirmed by Monte Carlo simulations
and cluster-variation calculations, see Ref.\ \onlinecite{DB}.
Therefore, we will always assume
that $|h|<2$ in the following, so that $T<T_c$ ($T>T_c$) corresponds
to the ordered (disordered) bulk phase.
}
\bibitem{def-OP}
{
As we shall see in Sec.\ \ref{GL}, a slightly different definition
of the local OP is favorable in order to perform the continuum limit
of the lattice MF equations. However, this alternate definition
changes none of the subsequent conclusions.
}
\bibitem{xi}
{
The lengths $\xi^\prime$ and $\xi$ [Eqs.\ (\ref{profiles100}a)
and (\ref{profiles110}a)] differ by a factor of $\sqrt{2}$ (to leading order
as $t\to0$) since we implicitly used the layer spacing as a unit length.
Therefore the bulk correlation length is $\sqrt{2}$ times larger
when measured in units of the (100) rather than the (110) layer spacing.
}
\bibitem{numerics}
{
Since the fixed point corresponding to the thermodynamically stable bulk phase
is unstable under the nonlinear mapping, any trajectory
computed numerically will eventually flow away from the fixed point, but the
error involved when calculating, e.~g., order parameter profiles can be made
arbitrarily small using standard techniques.
}
\bibitem{ampl-normal}
{
Beyond MF theory, the amplitude ratio $\phi_{s,+}^{(2)}/\phi_{s,-}^{(2)}$
of the $|t|^{2-\alpha}$ singularity of $\phi_s$ is known to be identical
to the universal value $B_+/B_-$, where $B_+$ and $B_-$ are the amplitudes
of the $|t|^{2-\alpha}$ singularity of the bulk free energy
(Ref.\ \onlinecite{normal}).
However, in MF theory one cannot separate the singular contributions
in the expansion of $\phi_s$ from the regular background terms.
Therefore, the ratio $\phi_{s,+}^{(2)}/\phi_{s,-}^{(2)}$
is a nonuniversal quantity in MF theory.
}
\bibitem{ampl-ord}
{
Without loss of generality we may assume that $h>0$, so that $g_1$ is
{\em positive\/} for $T<T_c$. Then the second solution of (\ref{surfEOS})
for $t<0$, which would lead to $\phi_{s,-}^{(1)}=g_1^{(1)}+A_1/\sqrt{B_0}$,
yields a higher surface free energy and may thus be discarded.
}
\end{references}
\end{document}